# From Forecast to Action: A Deep Learning Model for Predicting Power Outages During Tropical Cyclones

**Yongchuan Yang[a], Naiyu Wang[a,✉], Zhenguo Wang[b], Min Ouyang[c], and Can Wan[d]**

[a] College of Civil Engineering and Architecture, Zhejiang University, Hangzhou, China

[b] State Grid Zhejiang Electric Power Co., LTD. Electric power Research Institute, Hangzhou, China

[c] School of Artificial Intelligence and Automation, Huazhong University of Science & Technology, Wuhan, China

[d] College of Electrical Engineering, Zhejiang University, Hangzhou, China

✉ **Corresponding author**: naiyuwang@zju.edu.cn

**Abstract: Power outages caused by tropical cyclones (TCs) pose serious risks to electric power systems and the communities they serve. Accurate, high-resolution outage forecasting is essential for enabling both proactive mitigation planning and real-time emergency response. This study introduces the SpatioTemporal Outage ForeCAST (STO-CAST) model, a deep learning framework developed for real-time, regional-scale outage prediction during TC events with high-resolution outputs in both space and time. STO-CAST integrates static environmental and infrastructure attributes with dynamic meteorological and outage sequences using gated recurrent units (GRUs) and fully connected layers, and is trained via a Leave-One-Storm-Out (LOSO) cross-validation strategy along with holdout grid experiments to demonstrate its preliminary generalization capability to unseen storms and grids. The model produces hourly outage forecasts at a 4 km × 4 km resolution and supports dual forecasting modes: short-term nowcasting with a 6-hour lead time via assimilation of real-time observations, and long-term forecasting with a 60-hour lead time based on evolving meteorological projections. A case study on Typhoon Muifa (2022) demonstrates STO-CAST's operational effectiveness, including error decomposition across model design, meteorological uncertainty, and observation gaps, while highlighting the value of real-time data assimilation and the model's capacity to identify evolving outage hotspots. STO-CAST offers a scalable, data-driven solution to support risk-informed emergency response and enhance power system resilience under intensifying TC threats**.

## 1. Introduction

Tropical cyclones (TCs) are among the most destructive natural hazards, posing acute risks to coastal infrastructure systems—particularly electric power networks. As critical lifelines of modern society, power systems are highly susceptible to TC-induced disruptions, which can trigger widespread outages and cascading socioeconomic consequences (Hou et al., 2022). Recent events illustrate this vulnerability: Hurricane Ida (2021) left over one million customers in Louisiana without electricity for weeks (Holcombe et al., 2021); Typhoon Doksuri (2023) interrupted service to 1.9 million users in Fujian, China, necessitating the deployment of 45,000 emergency personnel (Dai, 2023); and Hurricane Helene (2024) caused more than 4.7 million outages across the southeastern United States, disabling over 100 transmission lines and 60 substations (Micek et al., 2024).

Unlike sudden-onset hazards such as earthquakes, TCs offer a valuable forecasting window that enables proactive risk reduction (Huang & Wang, 2024). Utilities can leverage this lead time—ranging from multi-day forecasts to short-term warnings—to implement protective measures such as reinforcing overhead structures (Ma et al., 2018; Panteli et al., 2017), elevating substations (Souto et al., 2022), pre-positioning mobile energy storage (Zhou et al., 2024) and issuing alerts to high-risk communities. In this context, accurate and timely outage forecasting functions as a critical intermediary—translating meteorological projections into actionable guidance that enhances both infrastructure resilience and emergency preparedness.

Modeling TC-induced outages remains a fundamentally challenging task due to the complex interplay of meteorological, geographic, and infrastructural factors. Outage dynamics are influenced by the evolving intensity and trajectory of the storm (e.g., wind, rainfall, and secondary hazards), heterogeneous terrain and land cover, and the physical vulnerabilities of aging infrastructure components (Hughes et al., 2022). Distribution networks are particularly prone to disruption, owing to limited redundancy and relatively lower design thresholds for critical assets such as poles, conductors, and transformers (Office, 2011; Salman et al., 2015; Zhai et al., 2021). A single failure—whether a downed line, collapsed pole, or substation malfunction—can cascade through the system, causing widespread downstream outages (Davidson et al., 2003). These interdependencies and the nonlinear propagation of disruptions significantly complicate the task of forecasting outages under TC conditions.

Over the past two decades, advances in data availability and computational methods have enabled significant progress in outage prediction modeling. Among these, **ensemble learning**—particularly tree-based methods—has gained widespread adoption for its robustness and interpretability. *Homogeneous* ensemble models such as Random Forest (RF) have consistently outperformed earlier statistical techniques (Nateghi et al., 2014), including generalized linear models (GLM) (Liu et al., 2005), generalized additive models (GAM) (Han et al., 2009), and hybrid regression approaches (Guikema & Quiring, 2012). For example, the Spatially-Generalized Hurricane Outage Prediction Model (SGHOPM) by Guikema et al. (2014) applied RF to forecast event-total outages over 3.66 km × 2.44 km grids using publicly available data. McRoberts et al. (2018) extended this work by introducing a two-step prediction procedure and incorporating additional predictor variables, some of which were previously explored by Quiring et al. (2011). Shashaani et al. (2018) developed a three-stage framework using RF to model zero-inflated event-total outages on 5 km × 5 km grids, demonstrating superior predictive accuracy by prioritizing a metric that penalizes false negatives. Additionally, Tonn et al. (2016) employed RF to analyze hourly outage data during Hurricane Isaac at the zip code level, revealing geographic variations in the statistical importance of storm-related covariates. More recent efforts have advanced toward *heterogeneous ensembles* that integrate multiple learners. The UConn Outage Prediction Model (Wanik et al., 2015) combined decision tree (DT), RF, and gradient boosted tree (BT) to predict event-total outages at 2-km resolution across 149 towns, with later expansions incorporating additional base learners (Cerrai et al., 2019) and extending coverage to five utility service territories (Watson et al., 2020). Hou et al. (2023) employed a stacking ensemble of extra tree, XGBoost, LightGBM, RF, and BT to estimate event-total outage durations at the 1-km scale under typhoon conditions. Furthermore, Kabir et al. (2024) proposed a feature- and performance-based weighting mechanism to adaptively combine outputs from multiple base learners to predict daily outages at the subarea level defined by utility companies.

Despite their strong spatial generalization performance, ensemble models are fundamentally static and primarily designed for event-total prediction rather than sequential forecasting. During training and inference, each input–output pair is treated independently, without modeling temporal dependencies. To approximate temporal variation, users often employ iterative input-updating methods (Guikema et al., 2014; Yang et al., 2021), reapplying the model at successive time steps with evolving or aggregated inputs (e.g., cumulative rainfall or maximum wind up to time *t*). Although this allows outputs to change over time, the model itself lacks an internal mechanism for learning temporal dynamics. Consequently, ensemble methods are inherently limited in capturing time-lagged relationships or forecast how outages cascading across consecutive time steps—constraints that might become particularly pronounced during fast-changing TC events.

To address these limitations, researchers have increasingly turned to **deep learning** approaches capable of modeling sequential dependencies. Recurrent neural networks (RNNs), particularly Long Short-Term Memory (LSTM) and Gated Recurrent Unit (GRU) architectures, have shown promise in learning temporal patterns within outage processes. For example, Alpay et al. (2020) used LSTM to forecast hourly thunderstorm-induced outages across 162 towns, while Abaas et al. (2022) developed an hourly LSTM outage prediction model for a single location. Tang et al. (2022) applied a GRU-based model to predict daily TC-related transmission line outages at the city scale. To strengthen spatial learning, hybrid models that combine convolutional neural networks (CNNs) with recurrent layers have also been explored. Udeh et al. (2022) integrated 1D CNN and LSTM for 48-hour predictions at county-level resolution; Huang et al. (2024) mapped meteorological features to simulated customer outage zones using a CNN-LSTM architecture; and Prieto-Godino et al. (2025) proposed a deep ensemble framework that blends RNN, LSTM, GRU, and CNN modules to predict province-scale daily outages.

Despite these advances, many deep learning models still operate at relatively coarse spatial resolutions—most commonly at the city or province level. This limitation is largely attributable to the scarcity of high-resolution, high-quality training data, particularly with respect to outage records and meteorological inputs. In cases where models do target finer spatial scales, they are often tailored to specific local contexts (e.g., individual substations or service zones), which constrains their generalizability to other regions with differing topographies, infrastructure characteristics, or hazard profiles.

Both ensemble and deep learning paradigms share two persistent limitations. *First* is the lack of simultaneous spatiotemporal resolution. Ensemble models often emphasize spatial accuracy but rely on aggregated temporal features and iterative updates to approximate dynamics—limiting their responsiveness to evolving storm conditions. Deep learning models effectively capture sequential patterns but often operate at coarse spatial resolutions, constrained by limited high-resolution training data. Few approaches achieve fine-grained resolution in both space and time—a capability critical for real-time, regional-scale, proactive emergency planning. *Second*, most models lack mechanisms to dynamically incorporate real-time meteorological updates or outage observations during inference. Some recent efforts have begun to explore adaptive elements, but their capacity remains limited in scope or application. For instance, Kabir et al. (2024) developed an adaptive ensemble model that uses real-time outage observation to enhance daily outage predictions. However, it does not currently incorporate evolving meteorological forecasts and operates at a temporal resolution less suited for intra-day emergency planning. Aljurbua et al. (2025) leveraged real-time social media and weather data to estimate short-term outage durations via a hierarchical spatiotemporal network, though their model excludes direct outage observations and exhibits reduced performance beyond a 12-hour horizon. These limitations underscore the value of rolling forecasting approaches that offer improved simultaneous spatiotemporal resolution by incorporating both forecasted and observed storm impacts at inference time.

Moving towards narrowing these gaps, we propose the SpatioTemporal Outage ForeCAST (STO-CAST) model—a deep learning framework tailored for high-resolution, regional-scale outage forecasting during TCs. STO-CAST augments existing methods by jointly processing static grid attributes and dynamic storm/outage sequences using gated recurrent units (GRUs) for time series and fully connected (FC) layers for spatial features. This architecture enables rolling forecasts that respond to newly available meteorological and outage inputs at inference time—without retraining or model updates—providing situationally relevant predictions throughout the storm lifecycle. Key contributions of the model include:

1) **Fine-Grained Spatiotemporal Forecasting**: STO-CAST generates hourly predictions over ~105,500 km² at 4 km × 4 km resolution, capturing the evolving disruption footprint throughout the storm lifecycle.
2) **Preliminary Cross-Event Adaptability**: Designed to learn from multiple TC events with diverse tracks and landscape features, STO-CAST demonstrates initial promise in generalizing to unseen conditions across four test storms—though further validation on additional events is warranted.
3) **Dual Forecasting Horizons for Real-Time and Proactive Response**: STO-CAST provides both short-term nowcasts (hourly forecasts up to 6 hours ahead) and longer-term forecasts (up to 60 hours). Short-term forecasts assimilate streaming outage observations to support immediate operational response, while longer-term forecasts leverage ~3-day meteorological predictions for pre-event planning and resource staging. This dual-horizon capability supports coordinated decisions across multiple phases of emergency operations.

By offering multi-horizon forecasts and real-time data assimilation, STO-CAST provides actionable insights for utilities, grid operators, and emergency management agencies. Its ability to capture evolving disruption footprints and adapt across multiple storms enhances situational awareness, supports dynamic resource deployment, and informs preemptive mitigation efforts. Recent advances in data availability suggest that rolling, high-resolution forecasting is increasingly feasible.

In the remainder of the paper, Section 2 details the data collection and preprocessing procedures utilized in the development of STO-CAST. Section 3 elaborates on the architecture and training process of the model. Section 4 presents a comprehensive evaluation of the model's performance. Section 5 applies the STO-CAST model to a case study of Typhoon Muifa (2022), demonstrating its practical utility and effectiveness in real-world scenarios. Finally, we conclude the study—highlight remaining limitations and future work—in Section 6. A schematic of the framework is presented in Fig. 1.

## 2. Data collection and preprocessing

Zhejiang Province, located in southeastern China, serves as the testbed to exemplify the general methodology for developing and implementing the STO-CAST model. Spanning 105,500 km² (Fig. 2), Zhejiang is one of the most densely populated and geographically diverse coastal regions in China. As of 2023, the province is home to approximately 66.27 million residents, distributed across 11 prefecture-level cities. The region is highly susceptible to TCs, experiencing an average of 2-3 events annually, which pose considerable challenges to its power infrastructure. For instance, Typhoon In-Fa (2021) caused outages affecting 2.68 million residents, while Typhoon Muifa (2022) disrupted power for approximately 520,000 households (Hu, 2022).

The STO-CAST model relies on two primary data categories: static and dynamic datasets. Static data include environmental characteristics, such as terrain and land use, and transformer exposure data, which provide foundational insights into the region's infrastructure vulnerability. Dynamic data encompasses meteorological observations and transformer outage records from specific TC events, capturing the evolving impact of storms on the power grid. Section 2.1 elaborates on the data sources, while Section 2.2 details the preprocessing procedures.

### 2.1 Data Collection

#### 2.1.1 Hazard Data: Meteorological Data and TC Track Information

Meteorological factors, such as strong winds and heavy rainfall, are primary drivers of power outages during TCs. High winds can damage power lines, while heavy rainfall may lead to flooding, submerging transformers, and causing short circuits. The track of a TC significantly influences the spatial distribution of wind and rainfall, making it another critical variable for characterizing TC

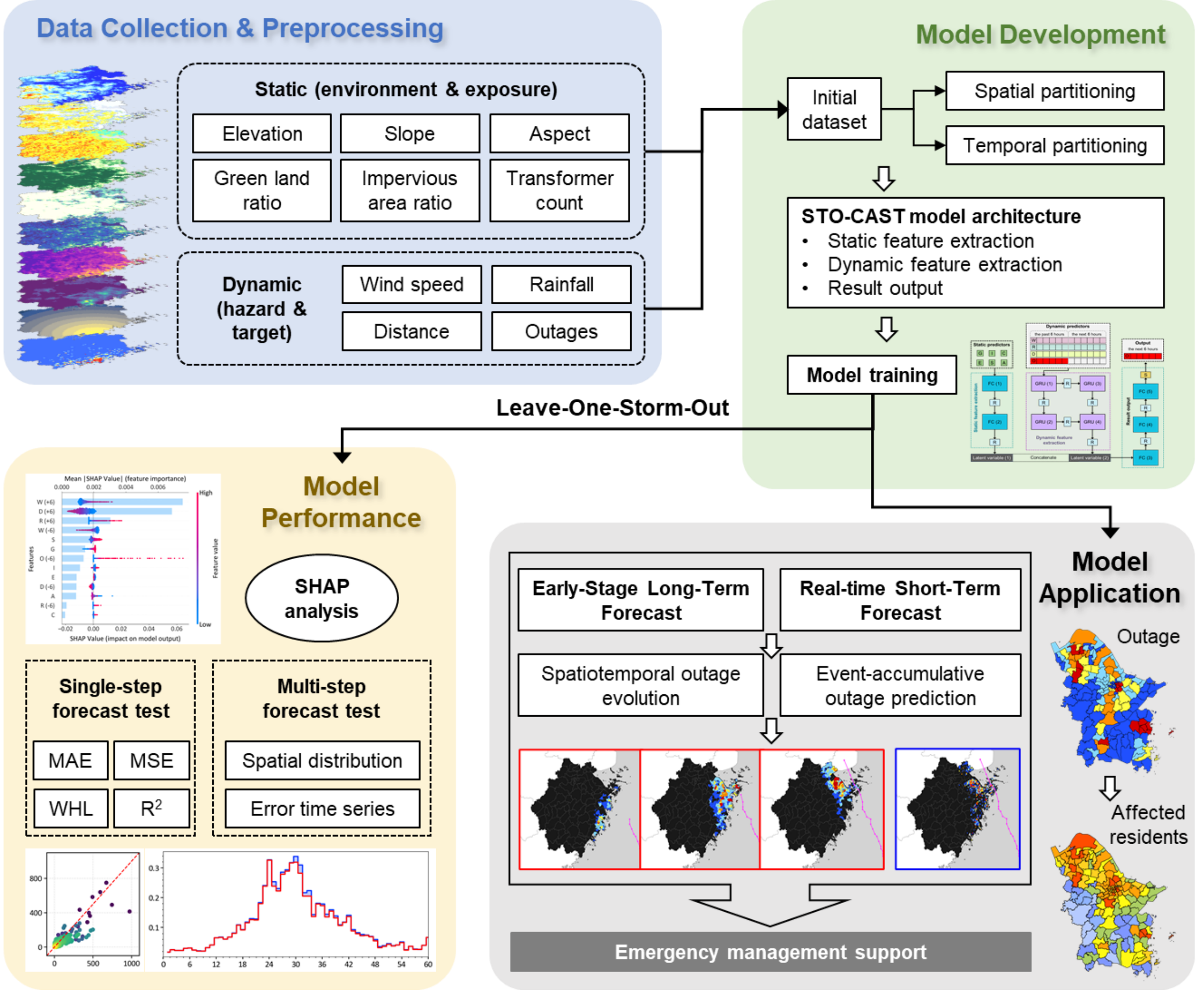

**Fig 1. Flowchart illustrating the development and application of the STO-CAST model**

dynamics. Together, these factors represent essential hazard variables for outage prediction.

High-resolution meteorological data, including hourly maximum 10-minute average wind speed (hereafter referred to as "wind speed") and hourly rainfall, are collected from 3,609 meteorological stations across Zhejiang Province for four TC events: Typhoons Lekima (2019), Hagupit (2020), Chanthu (2021) and Muifa (2022). These data are sourced from the Meteorological Bureau of Zhejiang Province, provide detailed meteorological observations. Additionally, TC track data, including the longitude and latitude of the cyclone eye recorded at 1-hour, 3-hour or 6-hour intervals, are obtained from the Oceanographic Data Center, Chinese Academy of Sciences (Wang, 2024). Table 1 summarizes the meteorological data collection periods, and Fig. 2 illustrates the tracks of these TCs.

### 2.1.2 Environmental Characteristics: Terrain and Land Use

Topography and land use significantly influence TC-induced wind speeds, rainfall distribution, and the severity of power grid damage. Variations in elevation influence wind intensity and precipitation levels (Miller et al., 2013), while land use patterns affect power infrastructure vulnerability (Petersen, 1982). Incorporating topographic and land use data is essential for improving the accuracy of outage predictions (Fatima et al., 2024).

Digital Elevation Model (DEM) data with a 30-meter resolution are sourced from ASTER GDEM (Abrams et al., 2020), while land use data with a 10-meter resolution are obtained from the Finer Resolution Observation and Monitoring of Global Land (Gong et al., 2019). Land use data categorize each pixel into one of ten classes: cropland, forest, grassland, shrubland, wetland, water, tundra, impervious surface, bare land, and snow/ice. These environmental characteristics, when combined with meteorological variables, enhance the accuracy of outage predictions.

### 2.1.3 Exposure Data: Transformer Distribution

Transformers, as critical components of power distribution networks, step down high-voltage electricity for end-users and are directly connected to consumers. Their failure directly impacts power supply reliability. Data for 213,291 transformers in Zhejiang Province, including geographic coordinates, are possessed by State Grid Zhejiang Electric Power Company. This dataset encompasses transformers managed by both public and private stakeholders and serves as a key predictor for representing vulnerabilities within the power distribution network.

### 2.1.4 Impact Data: Outages

Distribution transformers—hereafter “transformers”—are the low-voltage devices in the distribution network that step down power for delivery to end users (Tegene, 2023). In Zhejiang Province, each transformer is equipped with a Transformer Supervisory Terminal Unit (TTU) that continuously reports its operational status in real time. We define a "power outage" as any TTU-reported interruption in service, whether caused by transformer failure itself or by related infrastructure damage (e.g., downed poles or broken conductors). Because each transformer typically serves a small, localized customer group—such as several apartment buildings or a neighborhood—its outage record provides a reliable indicator of end-user impact.

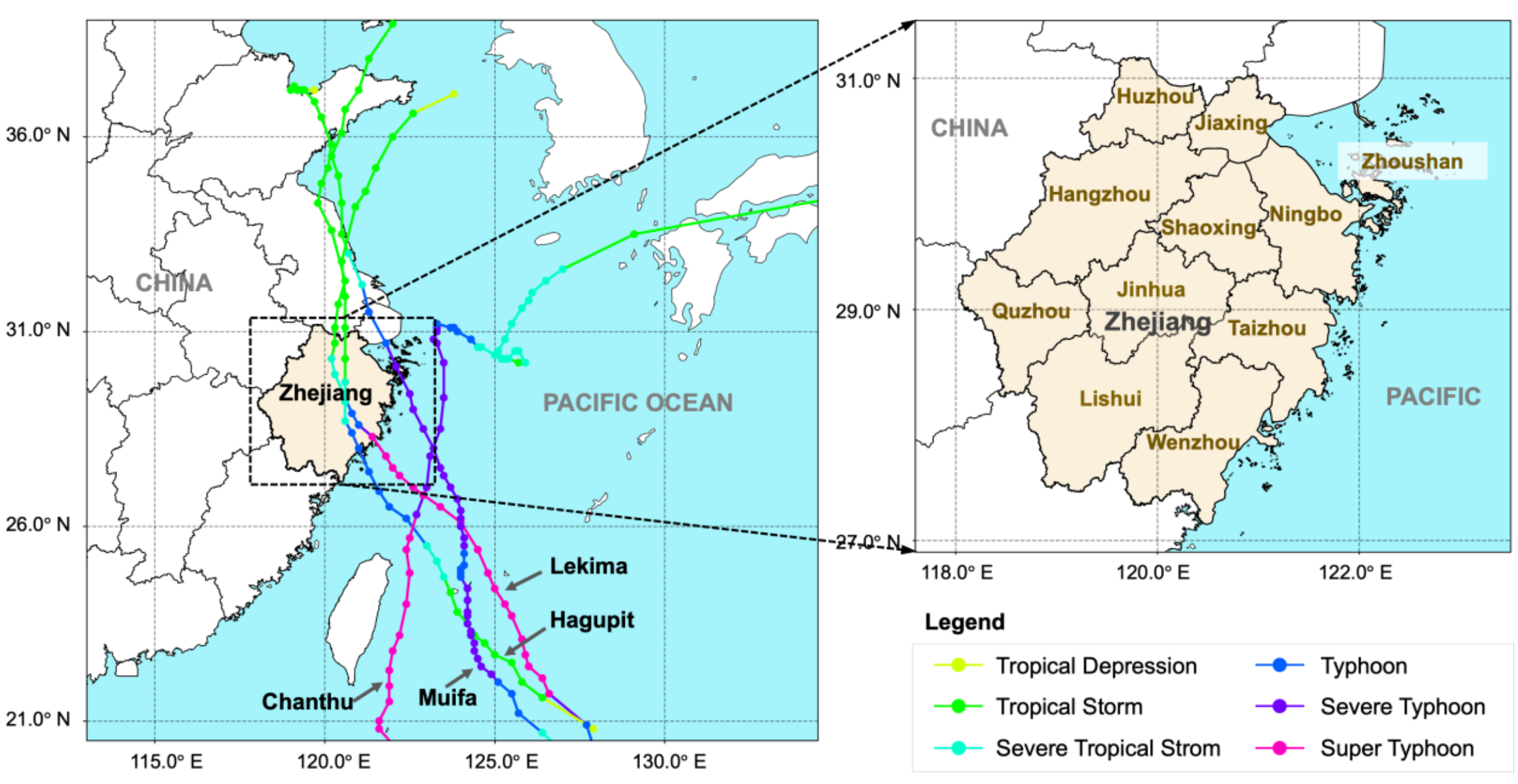

**Fig. 2 The location and administrate division of Zhejiang Province**

Outage data for the 213,291 transformers are collected across four TC events, including timestamps recorded at a 1-minute resolution. To improve data reliability, repeated outages occurring more than three times within one hour are excluded, as these are typically caused by signal disruptions during TCs. To align with meteorological data, outage timestamps are further standardized to a 1-hour resolution. For example, an outage recorded at 15:30 on August 10, 2019, is assigned to the 15:00-16:00 hour slot, assuming meteorological conditions during that hour was the influencing factor. Additionally, recurring outages at the same transformer during a TC event are consolidated, assuming that once an outage occurs, it remains in outage for the duration of the forecast period.

The outage count represents the total number of transformers experiencing outages in a given region, while the outage ratio is defined as the proportion of outage transformers relative to the total number of transformers in that region. As shown in Table 1, Typhoon Lekima (2019) caused the highest outage count in Zhejiang, while Typhoon Chanthu (2021), which did not make landfall, resulted in the fewest. Monitoring periods span approximately 36 hours before and 30 hours after landfall (or closest approach) for Typhoons Lekima, Hagupit, and Chanthu, and 30 hours before and 24 hours after landfall for Typhoon Muifa.

## 2.2 Data Partitioning and Assignments

### 2.2.1. Static Data

The grid size is a critical parameter that mediates the trade-off between spatial resolution and prediction accuracy. Smaller grids may lead to data imbalance and misalignment between outage and damage locations (Nateghi et al., 2014), while larger grids may lack the granularity needed to precisely identify outage hotspots. To optimize both predictive performance and computational efficiency, this study follows established best practices in geospatial modeling (Shashaani et al., 2018) by dividing Zhejiang Province (105,000 km²) into 4 km × 4 km grids, yielding 6,142 grid cells. This resolution is comparable to or smaller than the average town size—90% of Zhejiang's 1,346 towns exceed 16 km²—supporting localized outage prediction and facilitating town-level mitigation efforts.

Environmental variables, including elevation, slope, and aspect, are extracted from DEM data, with mean values calculated for each grid cell. Land use data are used to compute the impervious surface ratio and the green land ratio, the latter representing the proportion of the grid occupied by cropland, forest, grassland, or shrubland. These five environmental predictors—elevation (E), slope (S), aspect (A), impervious surface ratio (I), and green land ratio (G)—together characterize the topographic and ecological attributes of each grid cell.

The transformer count (C) within each grid cell is calculated as the sole exposure factor, ensuring simplicity while avoiding reliance on sensitive power network data. This parameter represents the critical components of power system and their associated vulnerabilities. The spatial distribution of these environmental and exposure predictors at the 4 km × 4 km resolution is illustrated in Fig. 3.

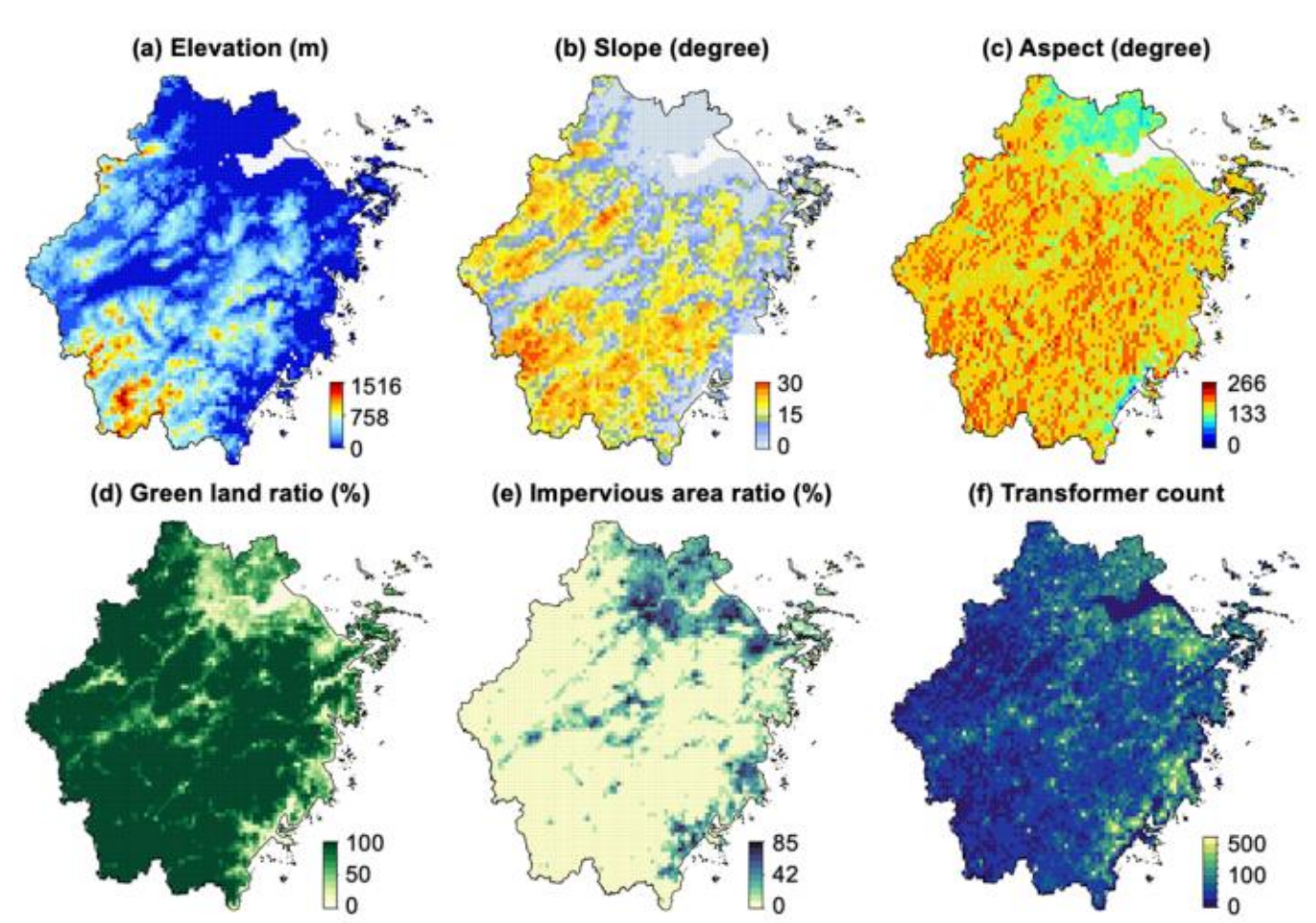

**Fig. 3 Environmental and exposure predictors**

### 2.2.2. Event-Based Dynamic Data

Dynamic variables, including meteorological data and outage data, are assigned to each grid to capture temporal variations during TC events. Wind speed and rainfall data from the 3,609 observation stations are interpolated spatially using the Inverse Distance Weighting (IDW) method, ensuring accurate representation of meteorological conditions. Each grid cell is assigned hourly sequences of wind speed (W) and rainfall (R) corresponding to the observation periods (c.f. Table 1). TC track data including the longitude and latitude of the cyclone eye, are standardized to a 1-hour temporal resolution using linear interpolation. This enables precise hourly calculations of the distance (D) from each grid centre to the TC eye.

**Table 1: Collection periods of event-based data**

| TC ID | Name | Year | Landfall time | Data start time | Data end time | Duration (hrs.) | Outage count |
|---|---|---|---|---|---|---|---|
| 201909 | Lekima | 2019 | 8/10 1:00 | 8/8 13:00 | 8/11 07:00 | 66 | 63,258 |
| 202004 | Hagupit | 2020 | 8/4 3:00 | 8/2 20:00 | 8/5 14:00 | 66 | 25,173 |
| 202114 | Chanthu | 2021 | - | 9/11 22:00 | 9/14 16:00 | 66 | 3,119 |
| 202212 | Muifa | 2022 | 9/14 20:00 | 9/13 15:00 | 9/15 21:00 | 54 | 9,057 |

We also map each outage to the transformer's location within a 4 km × 4 km grid. While this grid size may not pinpoint the exact fault on extended rural feeders, previous studies have demonstrated that 1–4 km aggregations remain robust for distribution system outage modeling (Hou et al., 2023; Yuan et al., 2020). We therefore assume that any spatial mismatch between the true fault location and the transformer's grid cell is negligible at this resolution. Nonetheless, we acknowledge that misallocated outages could bias feature attribution; future work will investigate fault-level modeling and incorporate higher-resolution location data to address this limitation. Outage data, initially recorded at 1-minute intervals, are aggregated to 1-hour resolution to align with meteorological data, as detailed in Section 2.1.4. The hourly outage counts per 4-km grid is calculated and assigned accordingly.

Fig. 4(a)-(d) illustrates the temporal evolution of the 4 km × 4 km grid-based wind speed, rainfall, minimum TC eye distance, and hourly incremental outage counts averaged across the entire Zhejiang Province during the 66-hour observation period of Typhoon Lekima (2019). Fig. 4(e)-(h) provides spatial snapshots of wind speed, rainfall, TC eye distance and accumulated outages across Zhejiang at 4 km × 4 km resolution at 10:00 on August 10, 2019.

### 2.2.3. Data Summary

After the data assignment process, each grid is characterized by ten variables, categorized into static and dynamic predictors, as well as a target variable:

1) Static Variables:
   - Environmental factors: elevation (E), slope (S), aspect (A), green land ratio (G), and impervious area ratio (I).
   - Exposure factor: transformer count (C);
2) Dynamic Variables:
   - Meteorological factors: wind speed (W), rainfall (R), and distance to the TC eye (D).
3) Target Variable:
   - Impact factor: transformer outage (O).

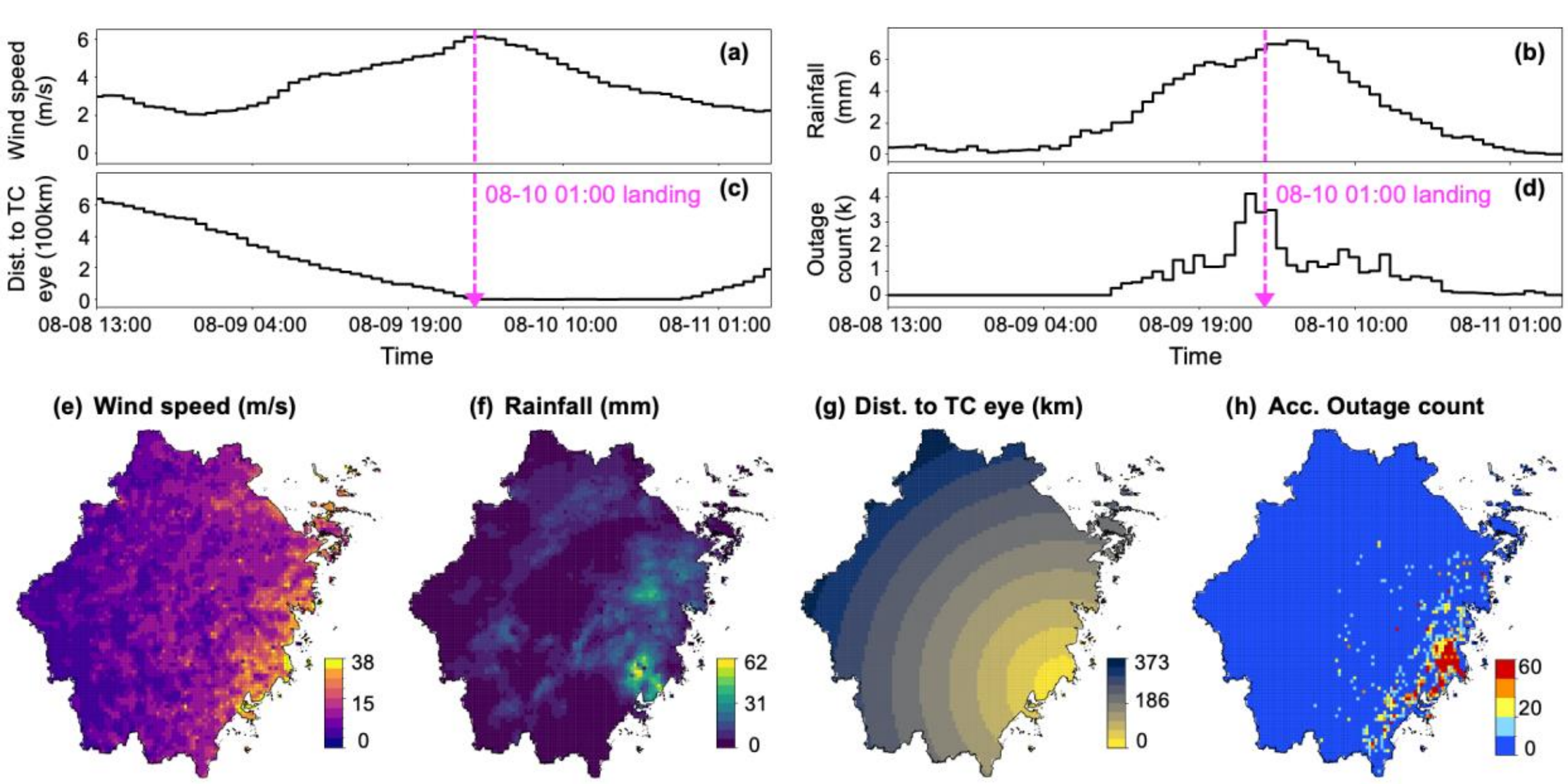

**Fig. 4 Temporal evolution (a-d) and spatial snapshots (e-h) of dynamic variables during Typhoon Lekima**

These predictors form the input features for the STO-CAST model, with outages (O) serving the prediction output. The dynamic data spans 66 hours for three TC events and 54 hours for one TC event, resulting in a total of 252 hours of data. Table 2 summarizes the dataset profile.

## 3. Model Development

### 3.1 Model Architecture

The STO-CAST model is deigned to efficiently process both static and dynamic predictors to forecast power outage time series during TCs. As illustrated in Fig. 5, the model takes as input four 12-hour time series, which include the past 6 hours (-6) and the next 6 hours (+6) of four dynamic variables: wind speed (W(-6) and W(+6)), rainfall (R(-6) and R(+6)), distance to the TC eye (D(-6) and D(+6)), and outages (O(-6)). For outages, the past 6 hours O(-6) are observed data, while the next 6 hours are zero padded to align with the other input sequences. Additionally, six static predictors — elevation (E), slope (S), aspect (A), green area ratio (G), impervious area ratio (I), and transformer count (C) — represent the environmental and exposure characteristics of each grid. By integrating both static and dynamic inputs, STO-CAST predicts the outage time series for the subsequent 6 hours O(+6).

The model architecture is built upon a combination of Gated Recurrent Units (GRUs) and Fully Connected (FC) layers. GRUs are particularly suited for sequential data processing, offering a simplified alternative to Long Short-Term Memory (LSTM) networks by mitigating vanishing and exploding gradient issues while maintaining lower computational costs (Chung et al., 2014). GRUs are employed to extract temporal features from the dynamic input sequences. FC layers are used to process the static predictors and latent variables for sequential prediction. The combination of GRUs and FC layers provides flexibility, effectively integrating static features and modeling the sequential dependencies for improved time series forecasting.

As illustrated in Fig.5, the STO-CAST architecture comprises three key modules: i) Static Feature Extraction Module: Processes the six static predictors (E, S, A, G, I, C) for each grid using two FC layers. ii) Dynamic Feature Extraction Module: Utilizes an encoder-predictor structure (Shi et al., 2015) with four GRU layers to process the four dynamic time sequences (W, R, D, O) over a 12-hour window, with the O(-6) time series being converted into outage ratio series. iii) Result Output Module: Concatenates latent variables from the static and dynamic modules at the channel level and passes them through three FC layers. The final FC layer applies a Sigmoid activation function (Mount, 2011) to generate the predicted outage ratio for the next 6 hours. This ratio is then multiplied by the transformer count in each grid to compute the predicted outage sequence O(+6). ReLu activation functions (Krizhevsky et al., 2012) are employed in the GRU and FC layers to ensure non-linearity, while a Sigmoid function for the final output layer. Table 3 summarizes the STO-

**Table 2: Initial dataset profile**

| Variables | Data Type | Data | Symbols | Temporal feature | Spatial resolution | Data Source |
|---|---|---|---|---|---|---|
| Predictors | Meteorological variables | Wind speed | W | 1h × 252 | 4km×4km | Wind and rainfall data observed hourly at 3,609 stations for the 4 TC events from Zhejiang Meteorological Agency |
| | | Rainfall | R | 1h × 252 | 4km×4km | |
| | | Distance to TC eye | D | 1h × 252 | 4km×4km | TC tracks from the Oceanographic Data Center, Chinese Academy of Sciences |
| | Geographic variables | Elevation | E | - | 4km×4km | 30m resolution digital elevation model (DEM) from ASTER GDEM (Abrams et al., 2020) |
| | | Slope | S | - | 4km×4km | |
| | | Aspect | A | - | 4km×4km | |
| | | Green land ratio | G | - | 4km×4km | Land use data at 10m resolution from the Finer Resolution Observation and Monitoring of Global Land (Gong et al., 2019) |
| | | Impervious area ratio | I | - | 4km×4km | |
| | Exposure variable | Transformer count | C | - | 4km×4km | Locations of transformers and outage observation data for 4 TC events collected by State Grid Zhejiang Electric Power Co., LTD. Electric power Research Institute |
| Target | Impact data | Outages (No. of transformers) | O | 1h × 252 | 4km×4km | |

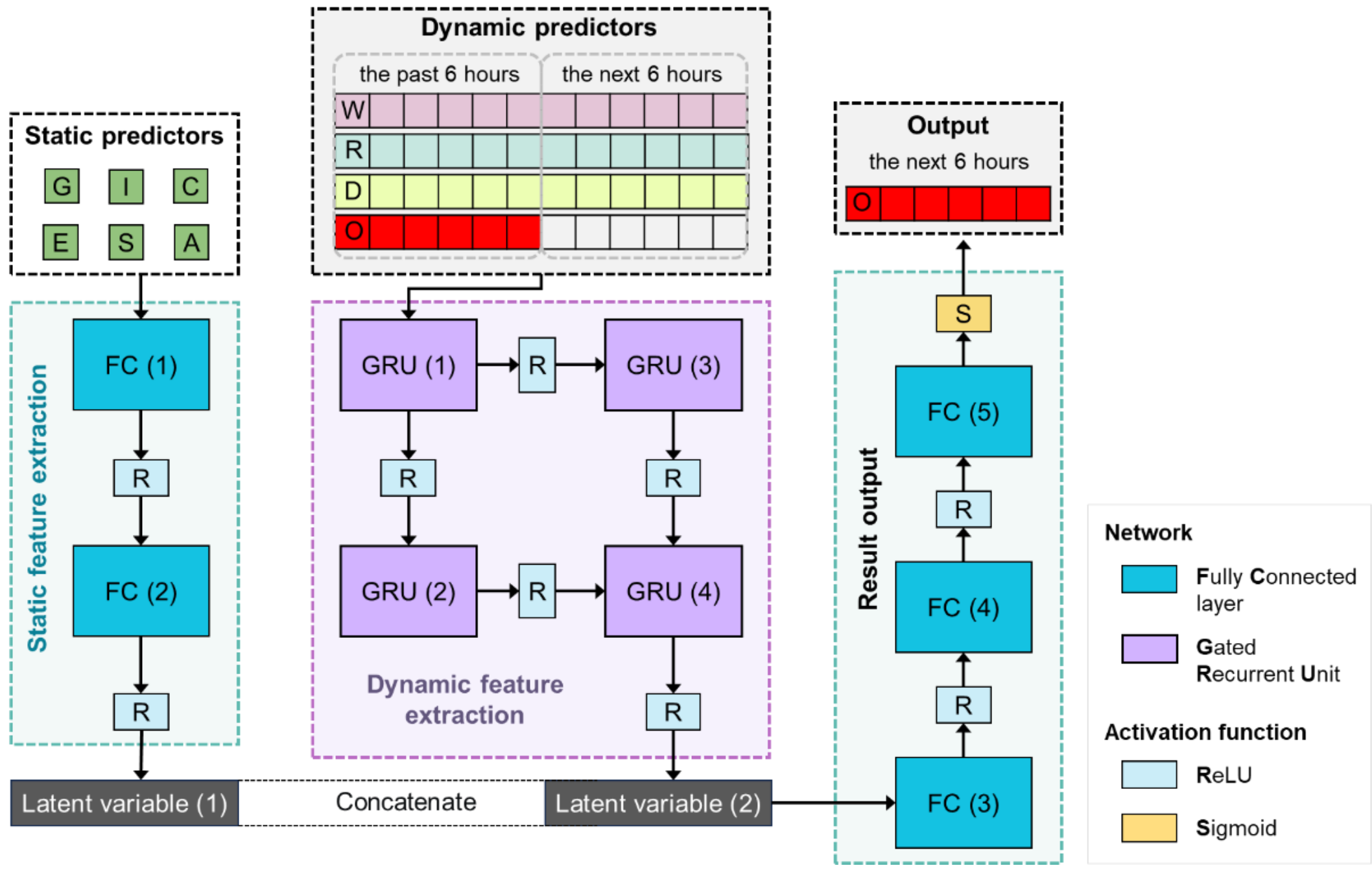

**Fig. 5 STO-CAST architecture and input-output**

CAST architecture parameters. Our focus is to demonstrate the feasibility of achieving improved simultaneous spatiotemporal resolution in outage forecasting. While incorporating additional spatial contexts or adopting alternative architectural enhancements could further improve accuracy, these directions are left for future work.

**Table 3: STO-CAST architecture parameters**

| Module | Network | Input channel | Output channel |
|---|---|---|---|
| Dynamic Feature Extraction | GRU (1) | 4 | 32 |
| | GRU (2) | 32 | 64 |
| | GRU (3) | 32 | 64 |
| | GRU (4) | 128 | 32 |
| Static Feature Extraction | FC (1) | 6 | 64 |
| | FC (2) | 64 | 32 |
| Result output | FC (3) | 224 | 64 |
| | FC (4) | 64 | 32 |
| | FC (5) | 32 | 6 |

## 3.2 Model Training

### 3.2.1 Dataset Partitioning

Due to the limited data from only four TC events, we adopt a Leave-One-Storm-Out (LOSO) cross-validation strategy (e.g., Alpay et al., (2020)) to evaluate the event generalization performance of the STO-CAST model. In each of the four experiments, one typhoon—Lekima (2019), Hagupit (2020), Chanthu (2021), or Muifa (2022)—is designated as the test event, while the model is trained solely on the remaining three. Each experiment is labeled after its hold-out test event (e.g., LOSO-Muifa uses Muifa as the test event). This design ensures that the test event is entirely unseen during training, enabling rigorous assessment of the model's cross-event generalization.

Within each LOSO experiment, the three training events' data are further partitioned spatially and temporally to enhance representativeness and prevent data leakage:

1. **Spatial Partitioning:** Of the 6,142 grids across the Zhejiang Province, 1,046 grids without transformers or observed outages are excluded from model training and testing. The remaining 5,096 grids are randomly divided into training and testing subsets with an 8:2 ratio (Fig. 6a). Specifically, 4,076 grids are allocated to the training grids, while 1,020 grids are reserved for testing. This partitioning ensured sufficient spatial diversity for model training and left unseen grids to evaluate the model's spatial generalization capability.
2. **Temporal Partitioning:** Dynamic data sequences for each grid are segmented using a sliding window approach. For example, each 66-hour time series of meteorological and outage data from a TC event is divided into overlapping 12-hour samples, sliding the window by 3-hour intervals (Fig. 6b). This approach results in 19 samples per grid per event, increasing the dataset size while capturing comprehensive temporal patterns. For outage time series, the next 6 hours of the input sequence are zero-padded, while the corresponding output labels are the original next 6 hours of the sequence (c.f. Fig. 5).

Taking LOSO-Muifa experiment as an example, the 3 training events (each spanning 66 hours) consist of 4,076 grids for training, with each grid contributes 57 overlapping 12-hour time sequences, resulting in a total of 232,332 training sequences. Of these, 185,865 sequences (80%) are random allocated for training (training set), and the left 46,467 (20%) for validation (validation set). The 1,020 grids reserved for testing contribute a total of 58,140 test sequences, referred to as test grids, which are utilized to assess the spatial generalization of the model. The Muifa event is withheld entirely during training and serves as the unseen test event in the LOSO-Muifa experiment for evaluating event-level generalization. All other LOSO experiments follow the same partitioning protocol.

This multi-dimensional partitioning strategy enables the model to learn from diverse spatiotemporal patterns while strictly separating test grids and test events—thereby supporting rigorous evaluation of both spatial and cross-event generalization capabilities.

### 3.2.2 Loss Function and Model Training

The dataset reveals significant class imbalance, with an average 6-hour outage ratio of ~0.1, indicating that outages occurred in average about 10% of grids at any given time. Additional challenges include noise and outliers caused by factors such as signal interference, human error, and cascading failures. To address these issues, a Weighted Huber Loss (WHL) function is employed, extending the traditional Huber Loss (HL) (Huber, 1964). HL combines the robustness of Mean Absolute Error (MAE) and Mean Squared Error (MSE), offering resilience to outliers. WHL further incorporates sample weighting to emphasize samples with high outage counts, ensuring the model pays greater attention to significant outage events.

The HL and WHL are defined as:

$$\mathrm{HL}_{\delta}\big(y, f(x)\big) = \begin{cases} \frac{1}{2}\big(y - f(x)\big)^2, & |y - f(x)| \le \delta \\ \delta|y - f(x)| - \frac{1}{2}\delta^2, & |y - f(x)| > \delta \end{cases} \quad (1)$$

$$\mathrm{WHL}_{w,t}\big(y, f(x)\big) = \begin{cases} \mathrm{HL}_{\delta}\big(y, f(x)\big), & y \le t \\ w \cdot \mathrm{HL}_{\delta}\big(y, f(x)\big), & y > t \end{cases} \quad (2)$$

in which $y$ is the observed outage value, $f(x)$ is the predicted outage value, $\delta$ controls the transition between MAE and MSE in HL, $w$ is the weighting parameter, and $t$ is the threshold for emphasizing specific samples. In this study, $\delta = 10$, $w = 1000$, and $t = 5$, ensuring that samples with significant outages are emphasized during training. This loss strategy ensures robust model performance despite the challenges of class imbalance and data noise.

Following Z-score standardization (i.e., subtracting the mean and dividing by the standard deviation) applied to the training set, the STO-CAST model is trained in PyTorch using the Adam optimizer (Kingma & Ba, 2014) with an initial learning rate of $5 \times 10^{-4}$. We employ a ReduceLROnPlateau scheduler that decreases the learning rate by 1% if the validation loss does not improve for 9 consecutive epochs. Training proceeds in batch size of 256 samples, and early stopping halts training after 99 epochs without validation-loss improvement. We save the checkpoint with the lowest validation loss as our final model.

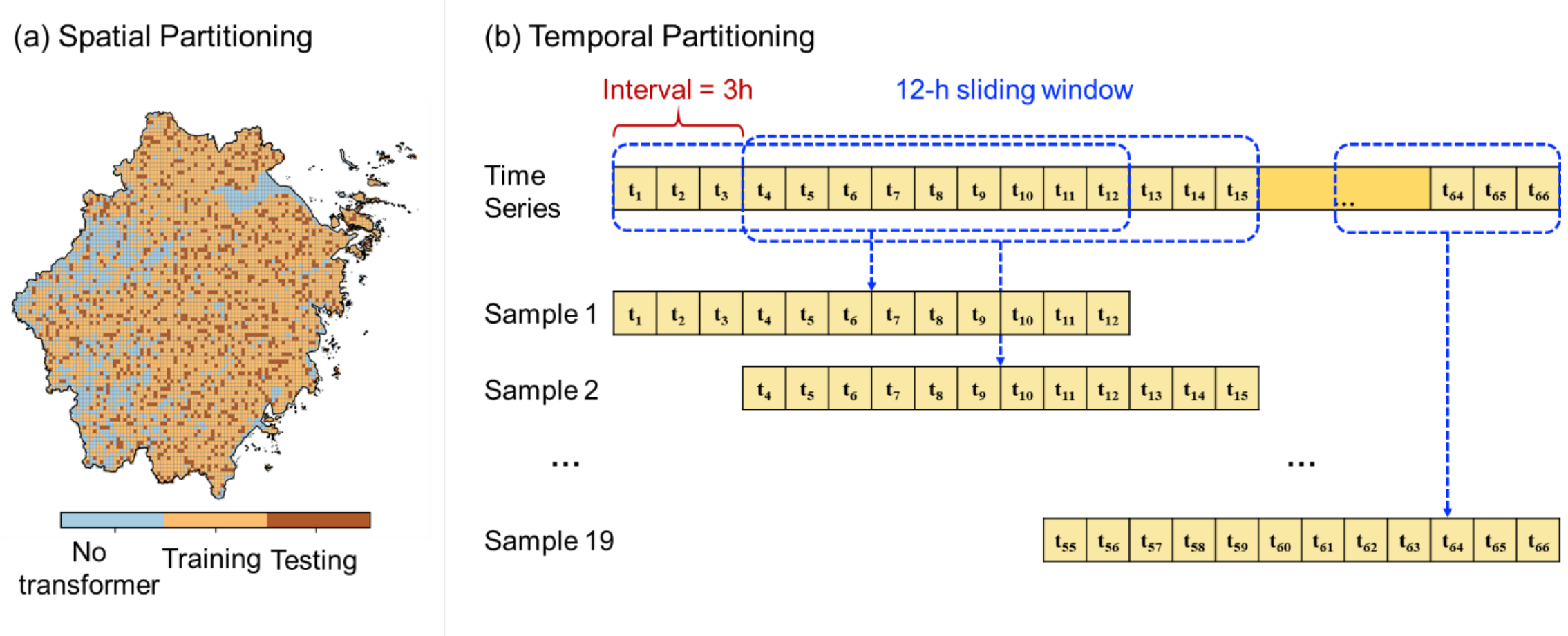

**Fig. 6 Spatial and temporal partitioning of the training data**

All experiments run on a machine with an Intel Core i7-13700K CPU and an NVIDIA GeForce RTX 4070 Ti GPU. Across the four LOSO experiments, the best model checkpoints occurred at epochs 185, 300, 239, and 209, respectively. Each training run requires approximately 35 minutes on this hardware setup.

## 4. Model Performance Evaluation

The performance of the STO-CAST model is systematically evaluated to assess its resistance to overfitting, generalizability to unseen data, and applicability across diverse forecasting scenarios. The evaluation includes three key components: an interpretability analysis using Shapley Additive Explanations (SHAP) (Lundberg & Lee, 2017), single-step prediction testing, and multi-step forecasting performance. Key findings from these analyses are presented below.

### 4.1 Interpretability Analysis

To illuminate the inner workings of the STO-CAST model, we apply the SHAP framework, which attributes predictive outcomes to individual input features based on cooperative game theory. Both static and dynamic predictors are assessed, with the dynamic features further split into past 6-hour histories (denoted as "−6") and forecasted next 6-hour inputs ("+6").

As an illustrative example, the LOSO-Muifa model is analyzed using 6,000 randomly selected 12-hour sequences from the training set as background data and 1,200 randomly selected sequences from the test-grid set for SHAP analysis. The results are visualized in Fig. 7, where horizontal blue bars summarize feature importance, and individual dots represent contributions for specific instances. The analysis reveals that forward-looking dynamic variables, such as the forecast wind speed W(+6), distance to the TC eye D(+6), and rainfall R(+6) for the next 6 hours, play the most significant role in determining the next 6-hour outage predictions O(+6). These predictors surpass the contributions of past dynamic variables like W(-6) and O(-6), underscoring the model's reliance on future trends for accurate forecasting. Static variables exhibit consistently low contributions, indicating their minimal influence on the predictive outcomes. Interestingly, the SHAP values for past outage observations O(-6) show notable variability across instances, suggesting that their impact is context-dependent and useful in certain cases. These findings align with the model's design, which prioritizes forward-looking predictors for accurate outage forecasting. The models trained in the other three LOSO experiments also yield similar SHAP results.

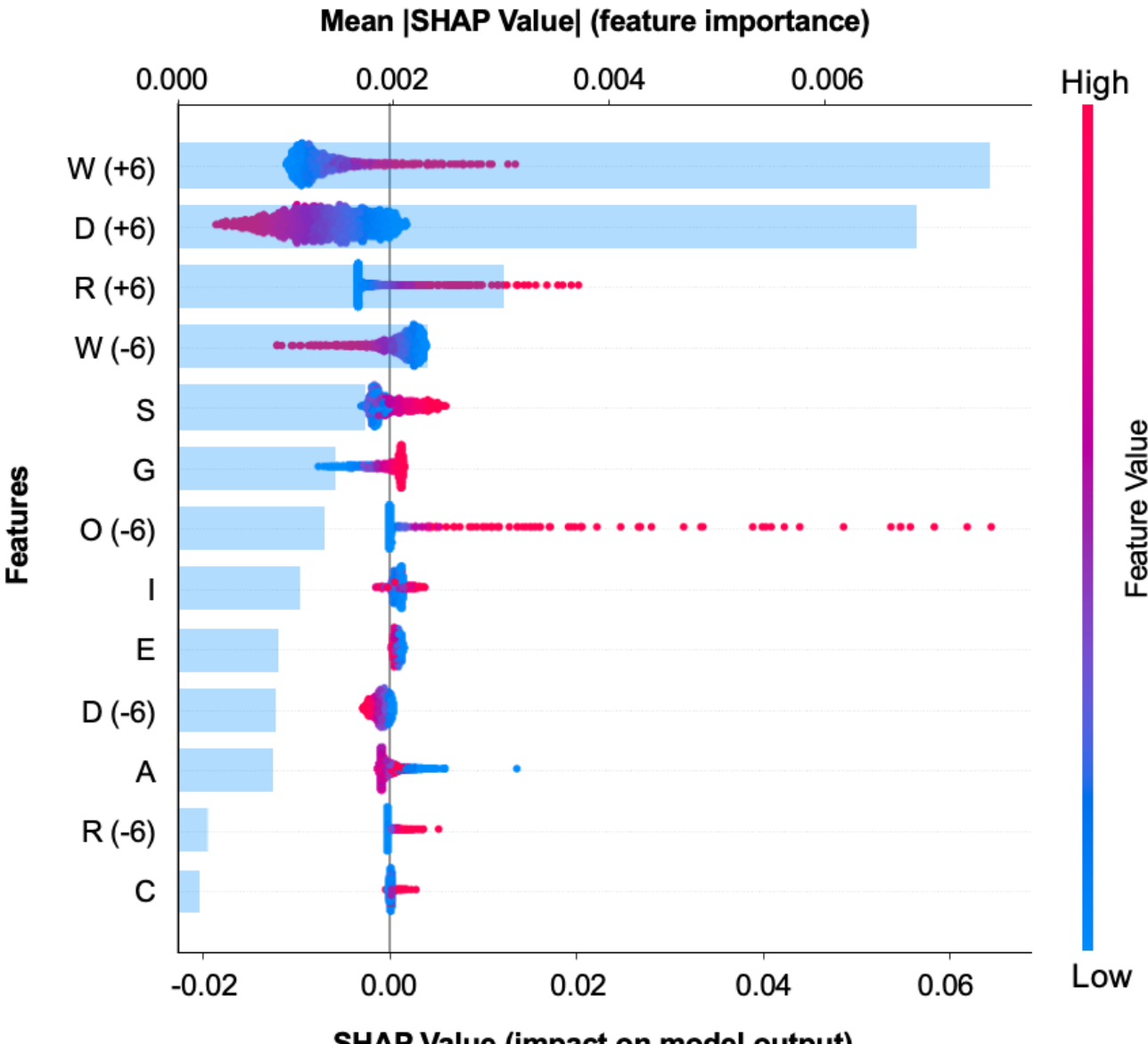

**Fig. 7 SHAP value analysis for feature importance**

### 4.2 Single-Step Prediction Test

To systematically evaluate the model's performance, we first assess STO-CAST's ability to generate 6-hour-ahead grid-based outage forecasts. Prediction performance is quantified using Mean Absolute Error (MAE), Mean Squared Error (MSE), and Weighted Huber Loss (WHL), computed across all datasets in four LOSO experiments.

As shown in Table 4, STO-CAST exhibits consistent performance across training, validation, and test-gird sets. The MSE and WHL metrics—which are particularly sensitive to rare but high-impact outages—remain comparable across different data splits (training vs. validation vs. test-grid sets), indicating minimal overfitting. Regression plots and $R^2$ statistics in Fig. 8 further support this observation. The predicted versus observed 6-hour cumulative outages show generally consistent alignment across data splits in four LOSO experiments, yielding an average $R^2$ of 0.66 for test-grid sets. Although performance is less accurate, this level of agreement suggests that STO-CAST can generalize reasonably well across unseen grids. These results provide an initial basis for future improvements aimed at increasing accuracy through expanded datasets and additional predictors.

**Table. 4 Performance metrics on grid-based 1-hour outage predictions**

| | LOSO-Lekima | | | LOSO-Hagupit | | | LOSO-Chanthu | | | LOSO-Muifa | | |
|---|---|---|---|---|---|---|---|---|---|---|---|---|
| | **MAE** | **MSE** | **WHL** | **MAE** | **MSE** | **WHL** | **MAE** | **MSE** | **WHL** | **MAE** | **MSE** | **WHL** |
| **Training set** | 0.050 | 0.408 | 3.935 | 0.100 | 0.761 | 14.87 | 0.134 | 1.115 | 28.23 | 0.120 | 1.066 | 25.35 |
| **Validation set** | 0.053 | 0.465 | 4.366 | 0.102 | 0.751 | 13.94 | 0.134 | 1.128 | 26.86 | 0.121 | 0.989 | 25.38 |
| **Test-grid set** | 0.050 | 0.585 | 3.695 | 0.108 | 0.917 | 16.54 | 0.136 | 1.288 | 26.28 | 0.120 | 0.972 | 23.37 |
| **LOSO test event** | 0.220 | 3.177 | 128.8 | 0.110 | 1.992 | 23.91 | 0.041 | 0.074 | 0.155 | 0.060 | 0.320 | 2.014 |
| (*Seen-event-baseline*) | (0.231) | (2.115) | (101.2) | (0.096) | (0.959) | (17.88) | (0.026) | (0.063) | (0.128) | (0.050) | (0.267) | (1.690) |

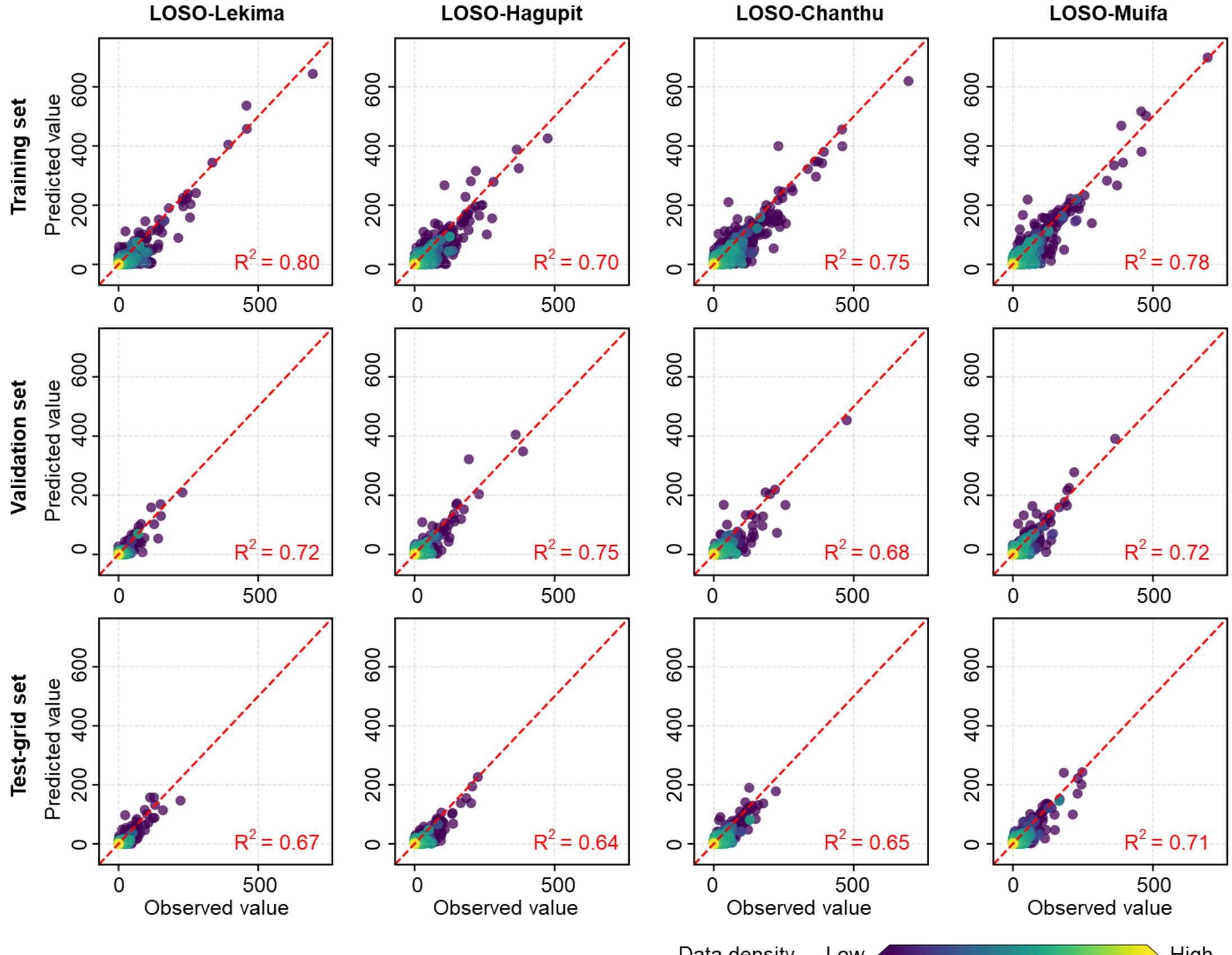

**Fig. 8 Four LOSO experiments' grid-based single-step 6-hour cumulative outage predictions vs. observations**

The final row of Table 4 ("seen-event baseline") presents the average performance on each test event when predicted by models trained with that event included. Due to data familiarity, STO-CAST performs better on the seen events. However, the performance of the LOSO results across four unseen events shows no significant difference in order of magnitude compared to the seen-event baseline, indicating that STO-CAST's generalization capability is not solely dependent on prior exposure to the test event. A preliminary evaluation of the model's event generalization performance is presented in Section 4.3. Nevertheless, we recognize that further improvements in event generalization are possible, and incorporating additional TC event data is expected to enhance this capability, a direction for future research.

## 4.3 Multi-Step Prediction Analysis

To evaluate the long-horizon forecasting and event generalization capabilities of STO-CAST, we conduct multi-step predictions across four LOSO test events, with each test event forecasted through multiple sequential 6-hour steps. For the Lekima, Hagupit, and Chanthu events, the forecasts consist of ten 6-hour iterations, covering a total of 60 hours per grid. For Muifa, eight iterations are used, spanning 48 hours.

The analysis is conducted under two distinct forecasting schemes—**Short-Term Nowcast** and **Long-Term Forecast**—demonstrating the model's flexibility in supporting both immediate response operations and early-stage proactive planning. In the **Short-Term Nowcast** scheme, the model assimilates updated observations of both outages and meteorological conditions every 6 hours. This iterative process ensures that each prediction step is informed by the most recent data, enabling responsive and adaptive forecasting with a rolling 6-hour lead time—an essential feature during the rapidly evolving phases of a storm. In contrast, the **Long-Term Forecast** scheme integrates outage observations and weather forecasts only once at the initial iteration. All subsequent predictions are generated by recursively propagating the model's own outage forecasts, using static weather inputs from the initial time step. As such, this scheme projects outage dynamics over an extended time horizon—up to the full span of the available weather forecast—without incorporating new observational data. While this approach sacrifices temporal adaptability, it is well suited for strategic pre-event planning, allowing decision-makers to allocate resources and bolster infrastructure resilience well in advance of TC impacts.

To isolate and rigorously evaluate the intrinsic predictive capability of the STO-CAST model, multi-step forecasting experiments are conducted under "*ideal weather conditions*", wherein observed meteorological variables are used in place of weather forecast inputs. This controlled setting eliminates the influence of weather forecast errors ($E_w$), allowing a focused assessment of the model's inherent predictive error, denoted as $E_{STOCAST}$ . Both the Short-Term Nowcast and Long-Term Forecast schemes are examined under this ideal condition to enable a direct comparison of their relative performance in modeling outage dynamics over extended time horizons—with and without the assimilation of real-time outage observations, respectively. Specifically, the prediction error of the **Short-Term Nowcast** under ideal conditions reflects only the model's inherent error, i.e., $E_{STOCAST}$; while the prediction error of the **Long-Term Forecast**, which does not incorporate updated outage observations beyond the initial step, includes both the model error and an outage observation input error, i.e., $E_{STOCAST}+E_o$. This evaluation provides a clean performance benchmark for the STO-CAST model, independent of weather forecast uncertainty.

As illustrated in Fig. 9(a), the multi-step prediction performance—measured by the MAE averaged across the four test events—shows minimal difference between the Short-Term

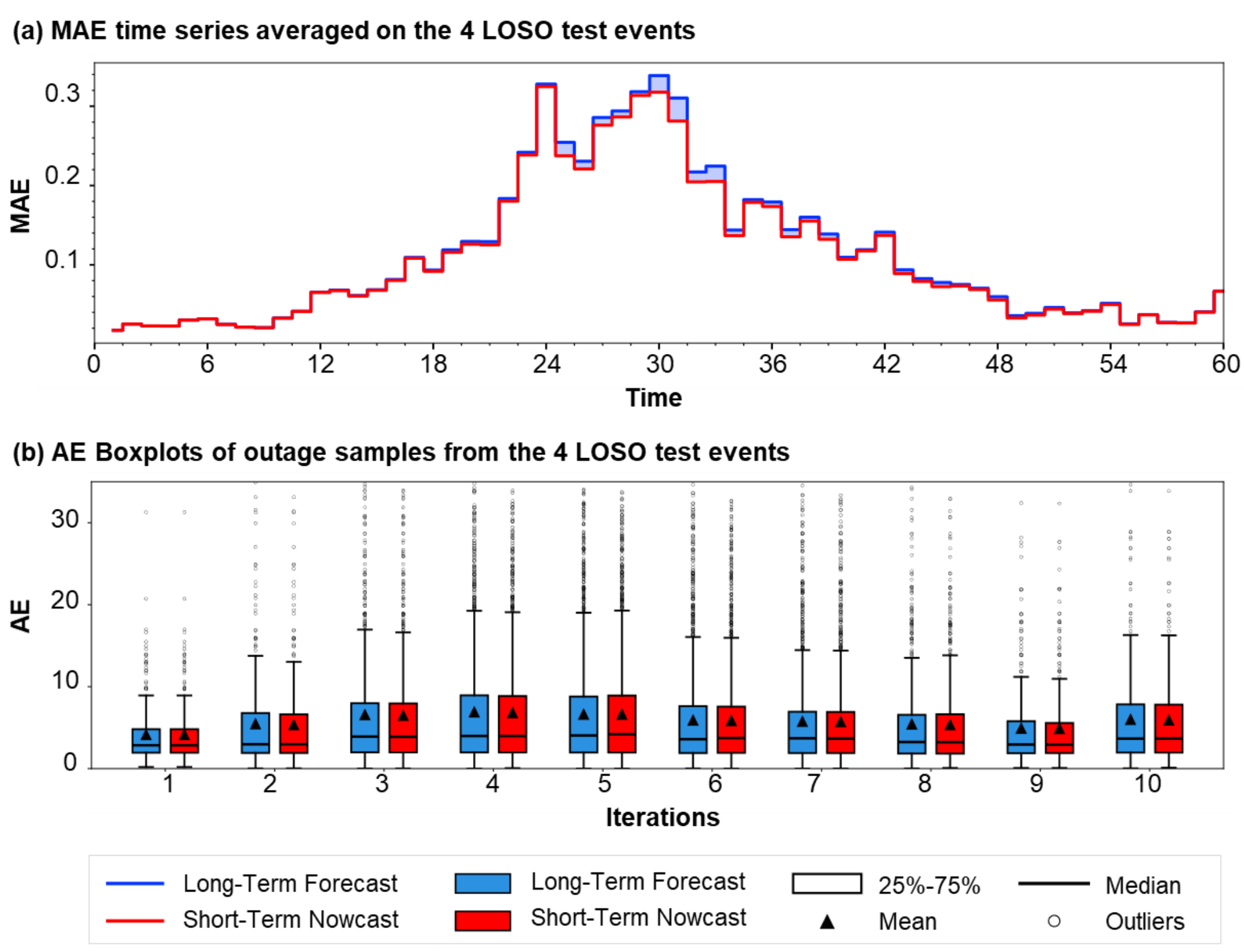

**Fig. 9 Grid-based multi-step error analysis across extended prediction period for four test events**

Nowcast and Long-Term Forecast schemes, particularly during the initial and final iterations. However, during intermediate time steps, which coincide with the periods of peak outage propagation, the Long-Term Forecast consistently exhibits marginally higher MAE. This discrepancy highlights the benefit of assimilating recent outage observations (i.e., O(-6)) in the Short-Term Nowcast, which modestly improves prediction accuracy during the most dynamically evolving phases of the event. These findings are aligned with the interpretability results in Section 4.1, where the STO-CAST model was shown to rely predominantly on forward-looking meteorological variables and only moderately on past outage inputs—especially during high-impact periods.

Additional evaluation is provided in Fig. 9(b), which presents boxplots of the AE calculated specifically for grid cells experiencing outages, thereby excluding the overwhelming number of zero-outage samples. This targeted assessment confirms that both forecasting schemes deliver consistently low prediction errors, with average AE values below 8 and median values below 5 outages per 6-hour step. Although slight increases are observed during peak periods, these remain tightly bounded, underscoring the model's robustness and stability in long-horizon, multi-step forecasting scenarios.

Fig. 10 assesses the STO-CAST model's ability to capture regional outage distributions under the Long-Term Forecast scheme across four LOSO test events. Panels (a) and (b) display the observed and predicted accumulative outages, respectively, revealing that STO-CAST effectively replicates regional-scale spatial patterns. When driven by ideal weather inputs, the model reasonably anticipates the location of major outage hotspots approximately 30 hours in advance of landfall, even though it modestly underestimates peak counts in some areas for Lekima and Hagupit. Panel (c) presents grid-level regression results, with $R^2$ values of 0.75, 0.64, and 0.72 for Lekima, Hagupit, and Muifa, respectively. For Chanthu—a non-landfall storm—slight overprediction is observed, likely due to the conservative bias induced by training on more intense landfalling events. Nevertheless, the predicted spatial patterns remain consistent with observations. The results from the four test events demonstrate the model's preliminary capability for generalizing across different events. Overall, these findings also highlight STO-CAST's potential utility for early-stage outage preparedness, offering both spatial reliability and actionable lead time for regional planning and resource deployment.

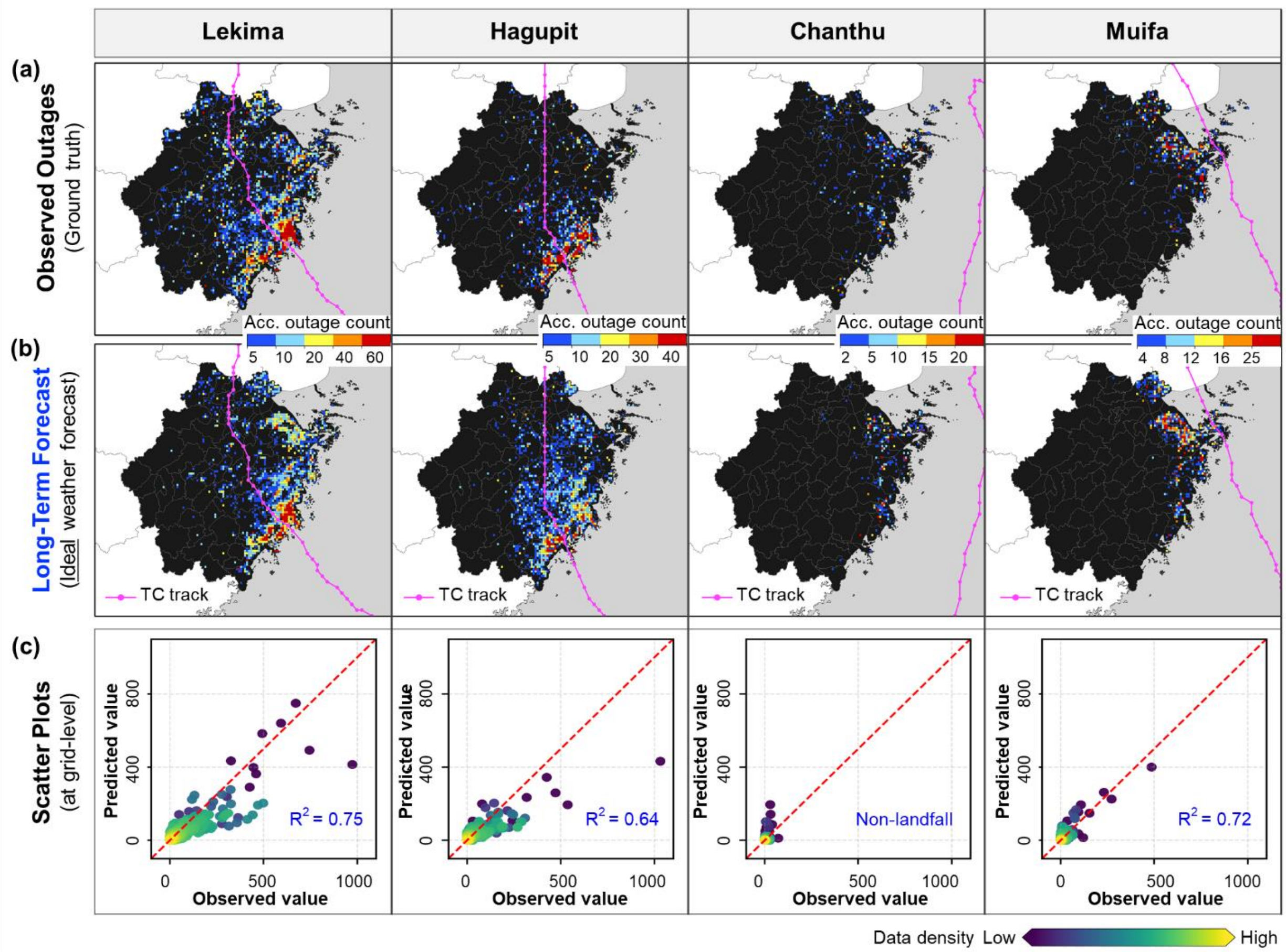

**Fig. 10 Spatial distribution of event-accumulative outages for the four LOSO test events: Lekima, Hagupit, Chanthu, and Muifa. Panel** (a) shows the observed accumulative outage counts; Panel (b) presents the corresponding forecasts generated by the LOSO model under the Long-Term Forecast scheme; Panel (c) displays scatter plots comparing predicted and observed grid-level accumulative outages, with associated $R^2$ values quantifying spatial agreement.

Fig. 11 assesses the real-time forecasting performance of STO-CAST for Typhoon Lekima using the LOSO-Lekima configuration under the Short-Term Nowcast scheme. Panel (a) compares observed and predicted accumulative outages at the provincial scale, with red triangles indicating 6-hourly data assimilation steps. Although the model slightly underpredicts outages prior to landfall, successive assimilation updates quickly improve alignment, suggesting that the mechanism is effective in adjusting forecasts based on incoming observations. Panels (b) and (c) present the spatial distribution of accumulative outages at four times: 7 and 1 hours before landfall, and 6 and 18 hours after, each with a 6-hour lead time. The observed patterns (Panel b) and model predictions (Panel c) exhibit broadly consistent spatial structures, particularly in identifying key outage hotspots. Although the model underestimates outage extent immediately prior to landfall (c-1 to c-2), it effectively captures the evolution and intensification of affected areas. From 6 hours post-landfall onward (c-3 to c-4), spatial agreement quickly improves and remains stable, supported by the model's ability to incorporate updated outage observations through data assimilation. Overall, these results indicate that STO-CAST can produce geographically resolved, short-term forecasts that respond to dynamic storm conditions—supporting timely decision-making in emergency response operations during TC events.

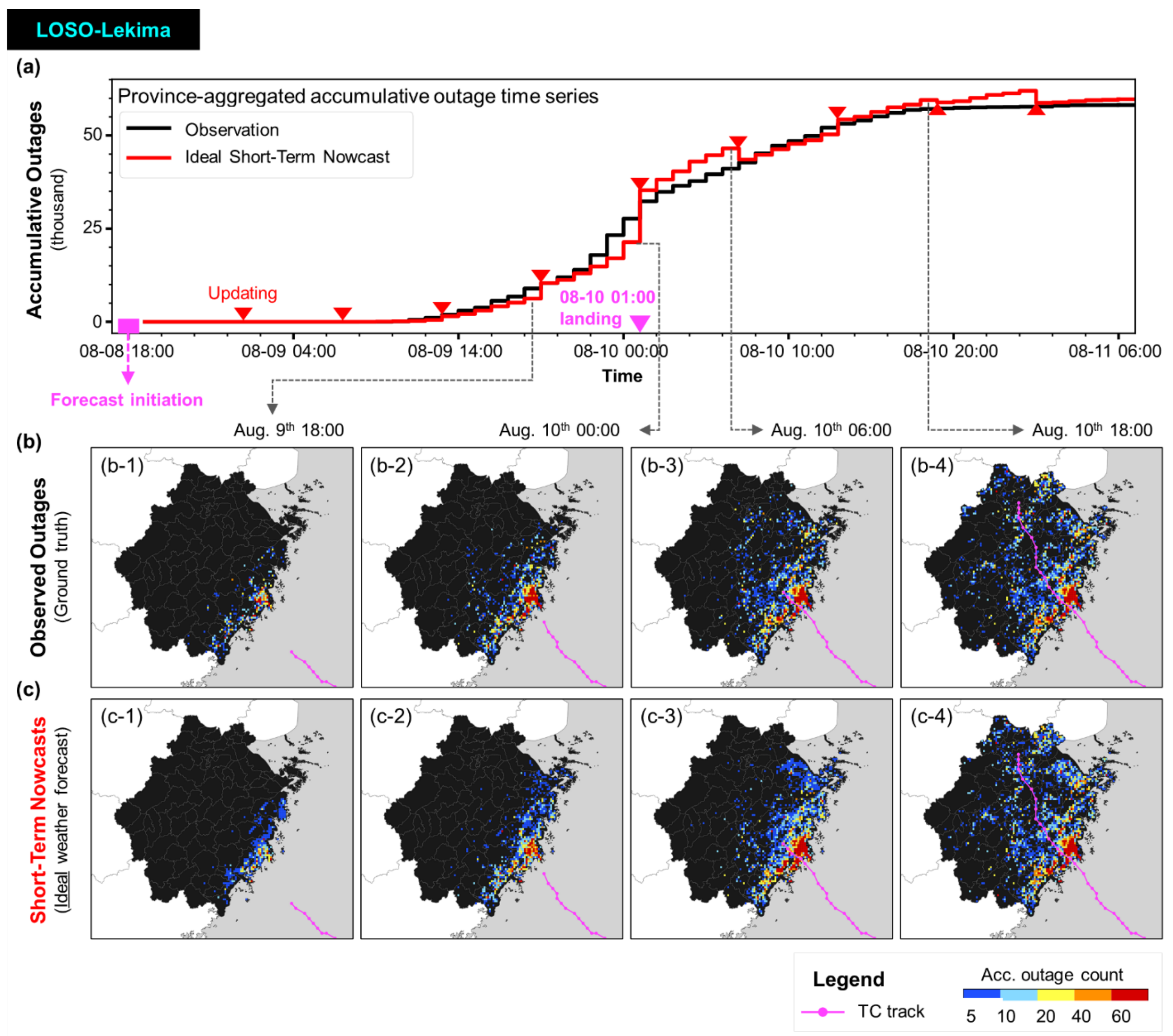

**Fig. 11 Performance of STO-CAST under the Short-Term Nowcast scheme for Typhoon Lekima (LOSO-Lekima).** (a) Provincial-level accumulative outages: observed vs. predicted, with red triangles marking 6-hourly data assimilation updates. (b) Observed and (c) Predicted spatial distribution of accumulative outages at four times: 7 h and 1 h before, and 6 h and 18 h after landfall.

It is important to emphasize that all results presented in **Section 4.3** are derived under *ideal weather conditions*, wherein observed weather variables are substituted for forecasted inputs. This controlled setting eliminates the influence of meteorological forecast errors, thereby allowing a focused evaluation of the intrinsic prediction error ($E_{STOCAST}$) and generalization capability of the STO-CAST model. However, in real event deployments, actual weather forecasts must be used, inevitably introducing additional errors—specifically, meteorological forecast uncertainty ($E_w$) and, in the case of the Long-Term Forecast scheme, compounding outage observation input errors ($E_o$) due to the lack of real-time assimilation. These compounded effects are analyzed in detail in Section 5, where the model's performance is evaluated under realistic operational conditions.

## 5. Model Application: Typhoon Muifa (2022)

The LOSO-Muifa configuration of the STO-CAST model was applied to predict power outages during Typhoon Muifa—an event entirely excluded from the training data—thereby providing a rigorous test of the model's event generalization capability and real-world utility. Typhoon Muifa made landfall in Zhoushan City, Zhejiang Province, at approximately 20:30 on September 14, 2022 (UTC+8), causing over 9,000 power outages across the region. To evaluate the model's utility, spatiotemporal outage predictions were generated over a 24-hour period starting at 1:00 AM on September 14—approximately 18 hours before landfall. Both forecasting schemes were employed: the Long-Term Forecast for early-stage mitigation planning and the Short-Term Nowcast for real-time response decision-making.

Meteorological forecasts (rainfall R and wind W) were provided by the Zhejiang Provincial Emergency Management Department, updated every 12 hours, with each forecast delivering a 24-hour outlook at a spatial resolution of 0.05 degrees (~5.5 km) and a temporal resolution of 1 hour. The congruent TC track forecasts (D), with a 6-hour temporal resolution, were obtained from the China Meteorological Administration (CMA) and interpolated to hourly intervals to align with meteorological data. Hourly outage observations (O(-)) were monitored in real-time by the State Grid Zhejiang Electric Power Company. Static input data (E, S, A, G, I, T) were sourced as outlined in Table 2.

Recall we have decomposed the model's total predictive error into three sources:

(1) $E_{STOCAST}$: model-intrinsic error, arising from the intrinsic limitations of the STO-CAST model, even when supplied with ideal input data.

(2) $E_w$: weather forecast error, resulting from inaccuracies or deviations in the meteorological forecast inputs (e.g., wind, rainfall, TC track).

(3) $E_o$ : observation assimilation error, caused by not assimilating outage observation in real-time, limiting adaptive correction of unfolding events.

These components collectively define the total outage prediction error observed during operational deployment. The following case application aims to quantify these components under real-world conditions.

### 5.1 Early-Stage Long-Term Forecast

The Long-Term Forecast scheme was used to generate the 24-hour lead-time accumulative outage prediction. Fig. 12 compares the spatial distribution of the observed 24-hour outages with the predictions using both *actual* and *ideal* weather forecasts. The predicted 24-hour outages under ideal weather conditions (Fig. 12(b)), conducted retrospectively to assess the model's performance excluding the influence of weather forecast errors (i.e., $E_{STOCAST} + E_o$), closely matched the observation (Fig. 12(a)), achieving a grid-based $R^2$ value of 0.73 for the entire Zhejiang Province. This strong correlation highlights the model's ability to identify critical hotspot areas with 24-hour lead time, particularly in the severely impacted regions.

In contrast, outage predictions driven by actual weather forecasts (Fig.12(c)), which include $E_{STOCAST} + E_w + E_o$, tended to overestimate the spatial extent of outages when compared to observations (Fig. 12(a)). This overprediction was primarily caused by upstream meteorological forecast errors—specifically in TC track (D), rainfall (R), and wind (W). To better understand the influence of these errors, Figs. 13(a)-(c) present the grid-based MAE of the 24-hour lead-time forecasts for R, W, and D (1st forecast in blue lines), averaged across Zhejiang Province. These forecast inaccuracies contributed directly to the observed discrepancies between outage predictions under actual (Fig. 12(c)) and idealized weather conditions (Fig. 12(b)). This contribution is quantified in Fig. 13(d) (blue line), where weather-induced errors account for a substantial share of the total forecast error. A notable example occurs in Taizhou, where the weather forecast severely overestimated meteorological hazards that did not materialize—leading to clear overprediction of outages in that region (red box in Fig. 12(c)). Nonetheless, when predictions were aggregated to the town level (Fig. 12(d)) and city level (Fig. 12(e)), the model still successfully identified the most impacted administrative regions—Ningbo, Zhoushan, and Jiaxing (cf. Fig. 1)—with the exception of Taizhou, where upstream forecast error led to inflated predictions.

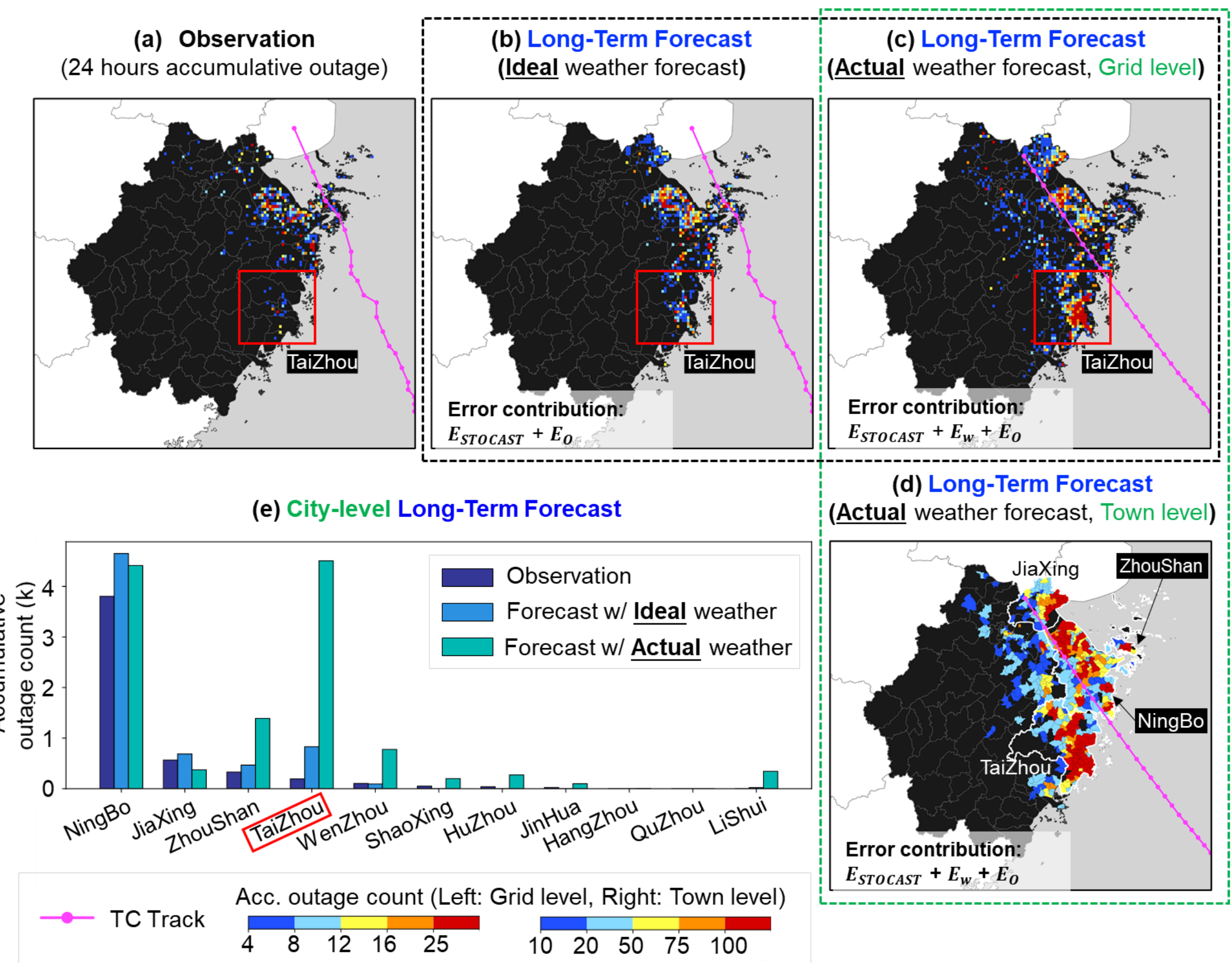

**Fig. 12 Observed and predicted 24-hour accumulative outages using Long-Term Forecast scheme**

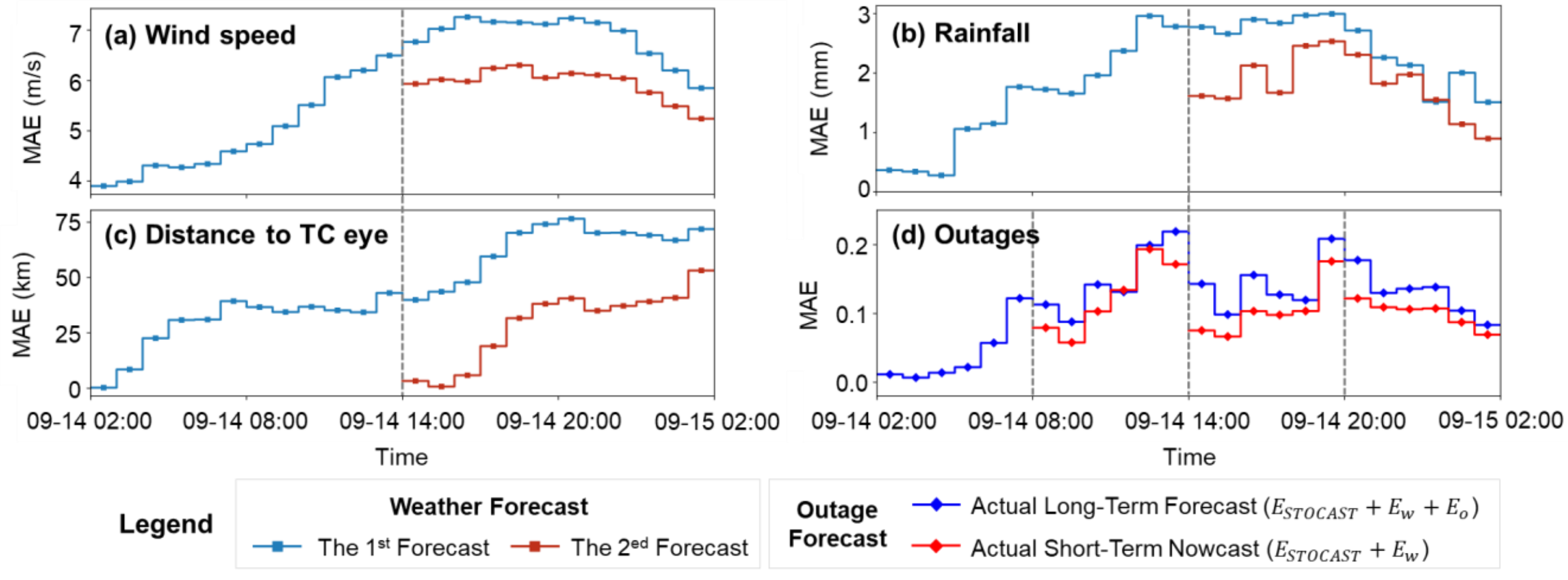

**Fig. 13 Time Series of MAEs for meteorological forecasts and outage forecasts**

Despite these discrepancies, the STO-CAST model's Long-Term Forecast under actual weather conditions effectively captured the overall outage trends, providing valuable 24-hour lead-time information for utilities and emergency response teams. While the prediction overestimated outages in Taizhou due to meteorological errors, this conservative bias ensures adequate preparedness for unexpected developments. As weather forecasting technology improves, this overestimation is expected to decrease, further enhancing the operability of the STO-CAST model.

## 5.2 Real-Time Short-Term Nowcast

**Fig. 14** evaluates the performance of the Short-Term Nowcast scheme in capturing the spatiotemporal evolution of power outages during Typhoon Muifa over a 24-hour prediction window, updated every 6 hours with a 6-hour lead time. This scheme incorporates the most recent information available— updated weather forecasts (wind (W+) and rainfall (R+), updated every 12 hours), revised TC track data (D+, updated every 12 hours), and newly observed outages (O-, updated every 6 hours)—to support real-time operational decisions.

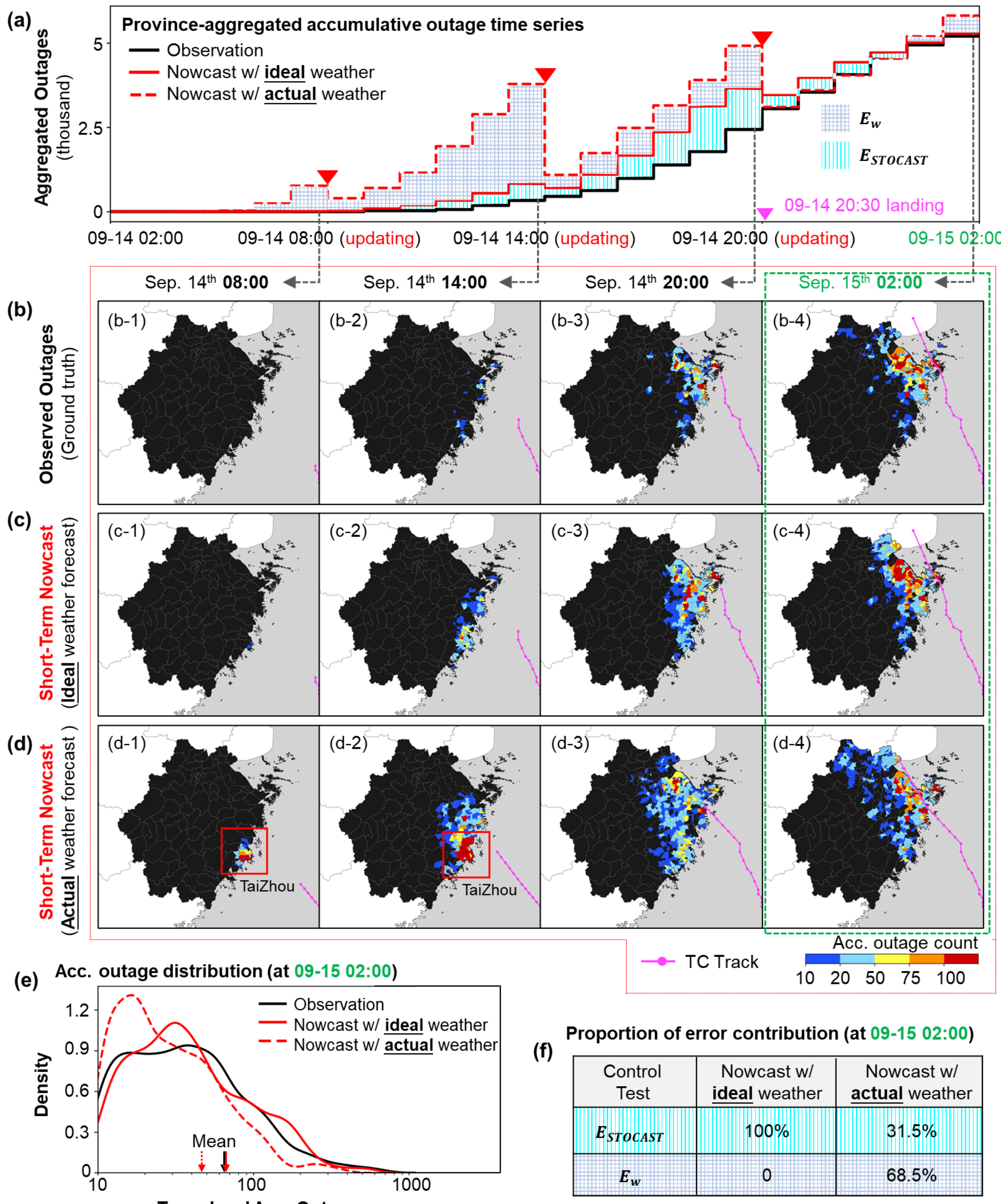

**Fig. 14 Spatial-temporal evolution of accumulative outages during Muifa (2022) (a)** Provincial-level accumulative outage time series under both ideal and actual weather inputs, **(b–d)** Spatial comparisons of observed and predicted accumulative outages at the town level, **(e)** Summary statistics at the 24-hour forecast horizon, and **(f)** A decomposition of prediction error into model vs. weather-induced components.

Panel (a) shows the time series of observed and predicted accumulative outages aggregated at the provincial level. Panel (b) maps the observed spatiotemporal evolution of town-level outages over the forecast horizon. Panels (c) and (d) present predicted outage patterns generated by the Short-Term Nowcast scheme under ideal and actual weather inputs, respectively. Under ideal weather conditions (Panel c), where prediction error arises solely from the model itself ($E_{STOCAST}$), the predictions closely track the observed progression of outage hotspots across the full 24-hour period (see comparison with Panel b and summary statistics in Panel a). In contrast, predictions based on actual weather forecasts (Panel d), which include both model error and weather-induced error ($E_{STOCAST} + E_w$), overestimate outages during the early hours—most notably in Taizhou—due to meteorological forecast inaccuracies, as discussed in Section 5.1. However, by incorporating updated meteorological inputs and newly observed outages (e.g., via the second assimilation step reflected in the red lines in Figs. 13(a)–(c)), the model adjusts its forecast in near real-time. Panels d-3 and d-4 show improved spatial correspondence with observations (Panels b-3 and b-4), including a marked correction of the initial overprediction in Taizhou.

Panel (e) provides a statistical comparison at the 24-hour mark (Sep 15, 02:00). Under ideal weather inputs, the model produces a mean accumulative outage prediction of 57.6 with a standard deviation of 65.3, closely aligning with observed values of 56.2 and 71.4. In contrast, predictions based on actual weather inputs yield a lower mean (38.9) and reduced variability (41.4), indicating degraded performance due to meteorological input uncertainty inherent in actual weather forecasts.

To further examine forecast reliability, Panel (f) decomposes the total prediction error. Under ideal conditions, the model-intrinsic error ($E_{STOCAST}$) accounts for 100% of the discrepancy. In contrast, under actual weather forecasts, the total error comprises both modeling error and weather-induced errors ($E_{STOCAST}+E_w$), with $E_w$—representing the deviation between predictions based on actual vs. ideal weather inputs—accounting for approximately 68.5% of total error (checkered area in Panel a), while the model error $E_{STOCAST}$ contributing only 31.5% (strip-filled area). This analysis highlights the model's inherent reliability and its strong dependence on meteorological forecast quality in operational settings. Accordingly, improvements in upstream weather predictions are expected to directly enhance the model's operational performance.

Finally, the Short-Term Nowcast consistently outperforms the Long-Term Forecast in predictive accuracy. This is illustrated by the red vs. blue trajectories in Fig. 13(d), and by comparing Fig.14(d-4) vs. Fig.12(d) against observed outages in Fig.14(b-4). The improvement stems from frequent data assimilation, which enables the model to adapt to rapidly evolving storm conditions. However, this advantage comes with the tradeoff of a shorter lead time—6 hours for the Nowcast compared to 24 hours for the Long-Term Forecast.

**Fig. 14** Spatial-temporal evolution of accumulative outages during Muifa (2022) **(a)** Provincial-level accumulative outage time series under both ideal and actual weather inputs, **(b–d)** Spatial comparisons of observed and predicted accumulative outages at the town level, **(e)** Summary statistics at the 24-hour forecast horizon, and **(f)** A decomposition of prediction error into model vs. weather-induced components.

### 5.3 Emergency Management

For emergency management at different phases, the Short-Term Nowcast scheme delivers accurate outage predictions with 6-hour lead-time, supporting real-time decision-making during the dynamic evolution of TCs. In contrast, the Long-Term Forecast scheme provides projections with an extended lead time, equipping utility operators to anticipate and strategically mitigate the potential impacts of TCs over a broader planning horizon. Together, the two complementary prediction schemes address both immediate response needs and strategic planning efforts, positioning the STO-CAST model as an effective tool for enhancing serviceability resilience of power utility before and during extreme TC events.

For example, Fig. 15(a) presents the observed and predicted accumulative outages aggregated at the city level during Typhoon Muifa in Ningbo—one of the most severely affected cities—using the Long-Term Forecast under ideal weather condition with a 24-hour lead time. The corresponding number of affected residents is calculated by overlaying outage forecasts with population data from the 2022 LandScan dataset (https://landscan.ornl.gov). The predicted trends closely matched observed data, offering a conservative yet reliable estimate of outages and their impacts. These insights enable power grid operators to anticipate large-scale outages and proactively adjust emergency response strategies.

Furthermore, aggregating grid-resolution outage and affected resident prediction to broader town- and county-levels, as shown in Figs. 15(b)-(i), enables the identification of towns and counties experiencing significant outages and substantial impacts on population, providing actionable insights at administrative resolution that is compatible to the scale of risk management decisions and actions. This information empowers power grid operators to optimize resource allocations among affected towns or counties, ensuring efficient and equitable risk mitigation operations.

## 6. Conclusions, Limitations and Future work

This study presents STO-CAST model, a deep learning-based framework for real-time, high-resolution power outage forecasting during tropical cyclones. By integrating spatiotemporal meteorological sequences, outage histories, and static infrastructure characteristics, the model provides hourly predictions at a 4 km × 4 km resolution across regional domains. Its dual-mode design—supporting Short-Term Nowcasts (6-hour lead time) and Long-Term Forecasts (up to 60-hour lead time)— enables both real-time operational response and pre-disaster planning. The model evaluation using a Leave-One-Storm-Out (LOSO) cross-validation strategy along with holdout grid experiments across four TC events, with results demonstrating its preliminary generalization capability to unseen storms and grids.

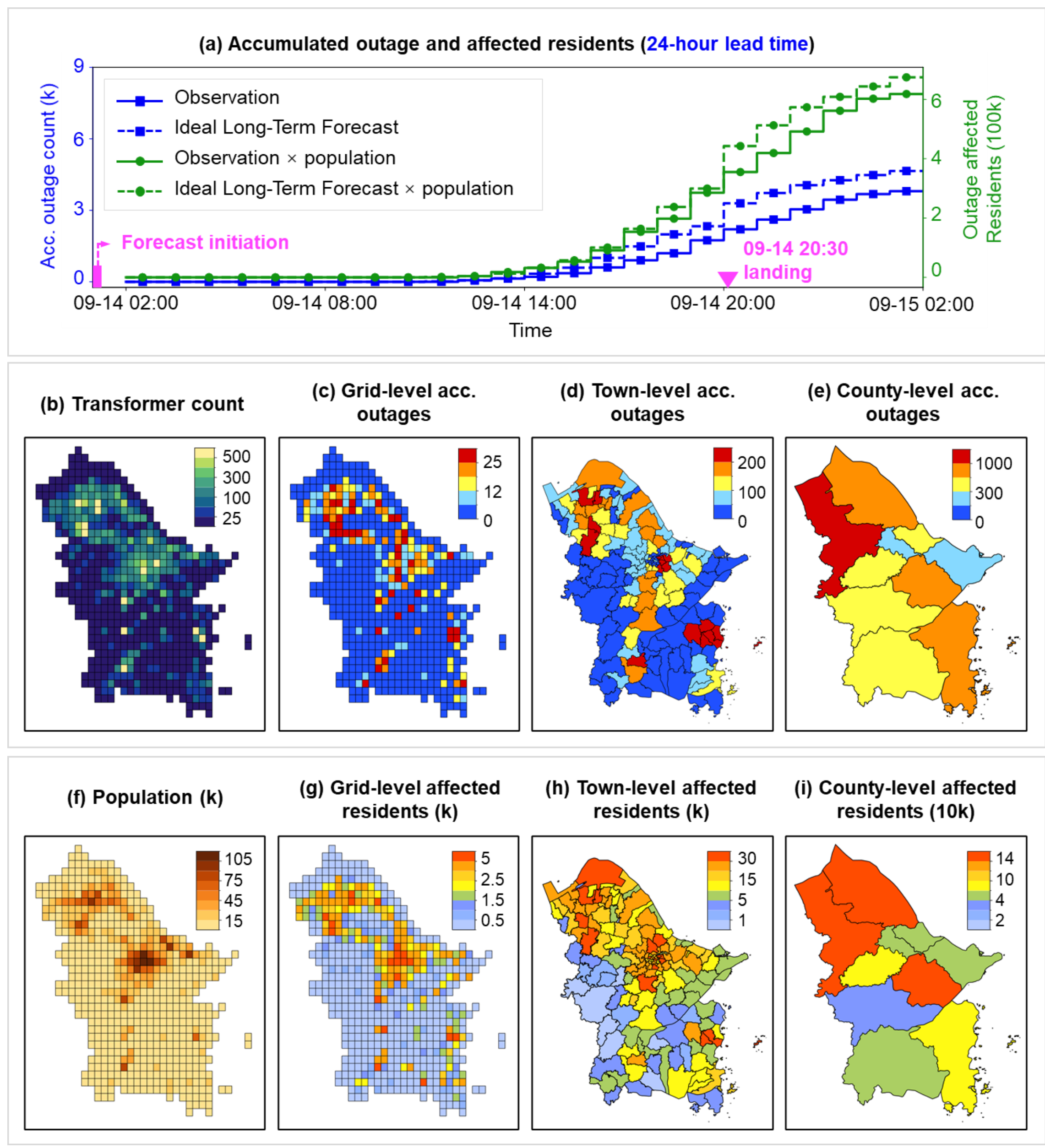

**Fig. 15** **Time series and spatial distribution of 24-hour accumulated outages and affected residents in Ningbo city**

Model testing and case applications show that STO-CAST reliably captures the spatial and temporal evolution of outages at fine granularity. The model benefits from assimilating both past and forecast inputs, adapting to evolving storm conditions and improving forecast accuracy, particularly during high-impact periods. Its performance under both ideal and actual meteorological conditions further clarifies the contributions of model error, meteorological forecast uncertainty, and observational data gaps to overall outage prediction accuracy. In particular, real-time assimilation of outage observations enhances short-term forecast accuracy, while long-term forecasts provide early situational awareness despite upstream uncertainties. These attributes support the model's practical utility for emergency interventions and proactive risk management.

Despite these strengths, the model has several limitations. The current training dataset comprises only four TC events, which constrains exposure to a broader range of storm characteristics. Outage data are recorded at the transformer level, which may not reflect the true fault origin—especially in rural distribution networks. Additionally, predictors from the physical distribution system remain limited, and model interpretability continues to pose challenges despite initial SHAP-based analysis. Further developments include expanding the training dataset and incorporating more detailed system-level features, refining the model architecture, integrating probabilistic meteorological ensembles, extending the framework to multi-hazard settings (e.g., wind-rain-flood impacts), and improving interpretability. These efforts aim to enhance the accuracy, generalizability, and transparency of STO-CAST, ultimately supporting more resilient and adaptive power system operations under evolving climate threats.

## Conflict of interest statement

The authors declare that they have no known competing financial interests or personal relationships that could have appeared to influence the work reported in this paper.

## Acknowledgements

This research was supported by the State Grid Zhejiang Electric Power Co., Ltd. Science and Technology Research Project (Grant No. B311DS24001A) and the Strategic Study Project of Chinese Academy of Engineering (2022-JB-02). This support is gratefully acknowledged.